\documentclass[useAMS,fleqn,usenatbib]{mnras}
\usepackage[T1]{fontenc}
\usepackage{ae,aecompl}
\usepackage{bethmacros}
\usepackage{graphicx}
\usepackage{amssymb}
\usepackage{amsmath}
\usepackage{hyperref}
\usepackage{tikz}
\usepackage{times}

\def\spose#1{\hbox to 0pt{#1\hss}}
\def\lta{\mathrel{\spose{\lower 3pt\hbox{$\mathchar"218$}}
     \raise 2.0pt\hbox{$\mathchar"13C$}}}
\def\gta{\mathrel{\spose{\lower 3pt\hbox{$\mathchar"218$}}
     \raise 2.0pt\hbox{$\mathchar"13E$}}}

\defcitealias{Kruijssen2014a}{KL14}
\title[Empirically-Motivated Feedback]{Empirically-motivated early feedback: momentum input by stellar
feedback in galaxy simulations inferred through observations}

\author[B.~W.~Keller, J.~M.~D.~Kruijssen and M.~Chevance]{Benjamin W.~Keller\thanks{Email: benjamin.keller `at'
uni-heidelberg.de},  J.~M.~Diederik Kruijssen and M\'elanie Chevance\\
Astronomisches Rechen-Institut, Zentrum f{\"u}r Astronomie der Universit\"at
Heidelberg, M{\"o}nchofstra{\ss}e 12-14, D-69120 Heidelberg, Germany}

\date{Accepted XXX. Received YYY; in original form 2021 September XX}

\pubyear{2021}

\begin{document}
\maketitle
\label{firstpage}
\pagerange{\pageref{firstpage}--\pageref{lastpage}}
\begin{abstract}
    We present a novel method for including the effects of early (pre-supernova)
    feedback in simulations of galaxy evolution.  Rather than building a model
    which attempts to match idealized, small-scale simulations or analytic
    approximations, we rely on direct observational measurements of the
    time-scales over which star-forming molecular clouds are disrupted by early
    feedback.  We combine observations of the spatial de-correlation between
    molecular gas and star formation tracers on $\sim100$~pc scales with an
    analytic framework for the expansion of feedback fronts driven by arbitrary
    sources or mechanisms, and use these to constrain the time-scale and
    momentum injection rate by early feedback.   This allows us to directly
    inform our model for feedback from these observations, sidestepping the
    complexity of multiple feedback mechanisms and their interaction below the
    resolution scale.  We demonstrate that this new model has significant
    effects on the spatial clustering of star formation, the structure of the
    ISM, and the driving of outflows from the galactic plane, while preserving
    the overall regulation of the galaxy-integrated star formation rate.  We
    find that this new feedback model results in galaxies that regulate star
    formation through the rapid disruption of star-forming clouds, rather than
    by highly efficient, global galactic outflows.  We also demonstrate that
    these results are robust to stochasticity, degraded numerical resolution,
    changes in the star formation model parameters, and variations in the single
    free model parameter that is unconstrained by observations.
\end{abstract}

\begin{keywords}
-- galaxies:formation --  galaxies:evolution -- galaxies:ISM -- 
    galaxies:star formation -- methods:numerical -- ISM:bubbles 
\end{keywords}

\section{Introduction}

Contemporary attempts to model the formation and evolution of galaxies have
found that the details of the involved feedback processes are perhaps the {\it
most} important component of both numerical and semi-analytic approaches
\citep{Somerville1999,Scannapieco2012,Rosdahl2017}.  Recent simulations of star
formation in giant molecular clouds (GMCs) and galaxies have included stellar
winds \citep{Rogers2013,Fierlinger2016,Wareing2017,Lancaster2021a}, direct radiation pressure
\citep{Krumholz2012,Raskutti2016,Costa2018}, supernovae
\citep{Keller2014,Kortgen2016,Grudic2019}, photoionization
\citep{Dale2017,Haid2019,Geen2021}, as well as different combinations of these
\citep{Agertz2013,Dale2014,Grudic2021}.  Despite these major efforts,
uncertainties in the energy losses and coupling efficiencies of these different
processes have led to a situation where in some cases, very different models
for stellar feedback are producing similar large-scale galaxy properties, while
in other cases models including the same physical processes are producing very
different galaxies.  Ideally, these processes could all be modelled from first
principles.  However, the computational cost of simulating the cosmological
evolution of Milky Way-like galaxies with sub-pc spatial and sub-$\Msun$ mass
resolution keeps this goal out of reach for the foreseeable future.   Instead,
we must continue to rely on approximations derived from analytic theory,
high-resolution simulations of individual molecular clouds, and observational
constraints.  Until now, applying observational constraints has been difficult,
and has been primarily relegated to post-hoc tests of simulated galaxies.

A major uncertainty in modelling stellar feedback in galaxy evolution is
quantifying how efficiently feedback energy couples to the interstellar medium
(ISM) to propel gas motions.  The momentum generation by feedback is critical
to disrupting GMCs \citep{Murray2010,Walch2012,Kruijssen2019b,Chevance2022},
regulating star formation \citep{McKee1977,Gatto2017}, driving turbulence in
the ISM \citep{Larson1981,Joung2006}, and launching galactic outflows
\citep{Lynds1963,Keller2015}.  Unfortunately, while it is simple to derive the
effect of a single feedback mechanism in toy models, the reality of multiple
feedback mechanisms operating simultaneously in fractal, non-spherical GMCs
greatly complicates determining how much feedback energy is converted into ISM
motions.  Even if the energy budget for feedback is well constrained, the
momentum budget is not.  Understanding how even a single feedback mechanism can
interact with a turbulent, multi-phase environment can require sophisticated,
ultra high-resolution simulations \citep{Fielding2020,Lancaster2021b}, which
are often poorly constrained by observations, and lack the full environmental
complexity of real star-forming GMCs.

With recent advances in the size and resolution of observational datasets,
along with new statistical tools to understand the implications of these
observations, it is now possible to directly constrain numerical models for
stellar feedback via observations. On kpc scales, galaxies follow a tight,
approximately linear relation between the (surface densities of the) gas mass
and star formation rate \citep[e.g.][]{Kennicutt1998,Bigiel2008,Leroy2008},
resulting in a roughly constant molecular gas depletion time (defined as the
ratio of the gas mass to the star formation rate; \citealt{Bigiel2011}).
\citet{Schruba2010} found that in M33 the depletion time of molecular gas
depends on the scale of the aperture over which these quantities are measured.
Measured depletion times diverge at smaller apertures depending on whether
those apertures are centred on gas or star formation peaks.  When an
observational aperture is decreased below the typical separation length between
independent star-forming regions $(\lambda)$, gas-centred apertures will mostly
contain clouds which have not yet been consumed by star formation or destroyed
by feedback.  Meanwhile, small apertures centred on peaks of star formation
tracers (H$\alpha$ for example) are more likely to sample regions where star
formation has progressed further, consuming (through star formation) and
disrupting (through feedback) dense gas.

The discovery of this de-correlation between gas and young stars on small
spatial scales motivated \citet{Kruijssen2014a} (\citetalias{Kruijssen2014a} in
further references) to develop an ``uncertainty principle for star formation'',
which is a statistical framework that uses the detailed shape of the spatial
de-correlation of depletion times as a function of aperture size (the ``tuning
fork'' diagram) to derive critical quantities for the timescale of star
formation and feedback.  With the method of \citetalias{Kruijssen2014a}, it is
now possible to systematically measure the lifetimes of molecular clouds
$(t_{\rm gas})$, the time-scale in which stars and gas are co-spatial $(t_{\rm
FB})$, the cloud-scale efficiency of star formation $(\epsilon_{\rm SF})$, the
separation length between independent regions of the ISM $(\lambda)$, and the
typical size of those clouds $(r_{\rm cl})$ directly from observations of star
formation and dense gas \citep{Kruijssen2018}.  With recent large, high
resolution, high sensitivity surveys such as LEGUS \citep{Calzetti2015}, PHANGS
\citep{Lee2022,Leroy2021}, and SIGNALS \citep{RousseauNepton2019}, it is now
possible to measure these quantitaties observationally across entire
populations of galaxies
\citep[e.g.][]{Kruijssen2019b,Chevance2020,Chevance2022,Ward2020,Zabel2020,Kim2021}.

In this paper, we will show how these observationally-measured quantities can
be used to build a new, empirically-motivated feedback (EMF) model that, by
construction, applies the observational star formation and feedback timescales
to determine the momentum injection rates of young massive stars.  Past
simulation studies \citep[e.g.][]{Fujimoto2019,Jeffreson2020,Semenov2021} have
applied the spatial de-correlation of gas and star formation as a post-hoc test
of feedback models.  These works have used observations of this de-correlation
to test whether their numerical models for star formation and feedback produce
realistic cloud lifetimes and feedback timescales.  EMF is the first stellar
feedback model that uses quantities measured through observations ($t_{\rm
FB}$, $\epsilon_{\rm SF}$, and $r_{\rm cl}$) as direct input parameters.

We structure the paper as follows.  In Section~\ref{FB_mechanisms}, we provide
a brief overview of the different mechanisms that can disrupt molecular clouds
through stellar feedback. In Section~\ref{momentum_derivation}, we show how to
derive the momentum injection rate and terminal momentum from quantities
measured through the \citetalias{Kruijssen2014a} method for different feedback
mechanisms.  This approach is used to build the EMF model, which we
subsequently implement in numerical hydrodynamical simulations of galaxy
evolution in Section~\ref{simulations}. There, we present the first simulation
results using this novel model for early stellar feedback, and show how this
early feedback can change the mode of star formation regulation, outflow
driving, and the structure of gas and stars in a Milky Way-like galaxy.  We
conclude with a comparison of our results to other approaches for including the
effects of early stellar feedback in Section~\ref{discussion}, and summarize
our main results in Section~\ref{conclusions}.

\section{Feedback Mechanisms for Disrupting Molecular Clouds}
\label{FB_mechanisms}
\subsection{Supernovae}
Supernovae (SNe) are one of the first feedback mechanisms proposed to explain
the hot galactic coronae \citep{Spitzer1956}, observations of outflows in
starburst galaxies \citep{Lynds1963}, and the paucity of gas in elliptical
galaxies \citep{Larson1974}.  A typical core-collapse SN will release
$\sim10^{51}\;\rm{erg}$ of energy after the death of a massive $(>5-10\Msun)$
star \citep{Ekstrom2012}.  This means SNe alone produce enough energy to unbind
typical Galactic-disc GMCs \citep{Reina-Campos2017}.  High-resolution
simulations of supernovae in molecular clouds indeed show that GMCs, after
preprocessing by stellar winds or radiation, can be completely disrupted by SN
feedback \citep{Rogers2013}.

The classic \citet{McKee1977} model of a three-phase ISM relies on a hot, volume
filling phase of SN-heated gas, in which cool atomic and molecular clouds
reside, enveloped by an intermediate warm ionized phase.  This is essentially
the picture which the methodology of \citetalias{Kruijssen2014a} examines.
\citetalias{Kruijssen2014a} provide a mechanism for determining the time-scale
in which the cool atomic/molecular clouds are disrupted, and a characteristic
separation scale of these cool clouds.

When it comes to translating the measurements made with the technique of
\citetalias{Kruijssen2014a} into a physical model of feedback in the ISM, core
collapse SNe (SNII) have the convenient feature of a built-in time-scale.  For a
given stellar population with a fully-sampled initial mass function (IMF), the
first SNII will detonate $\sim 3.5\Myr$ \citep{Maeder1989,Ekstrom2012} after the
formation of the most massive stars.  This means that if the observations yield
feedback time-scales shorter than this, we can rule out SNe as the mechanism for
destroying star-forming clouds \citep{Chevance2022}.  Uncertainty in this
time-scale due to incomplete IMF sampling \citep{Chabrier2003,Kroupa2003} or the
influence of binarity \citep{Eldridge2008} or rotation \citep{Leitherer2014} on
stellar lifetimes all push this time-scale up, so $3.5\Myr$ is really a minimum
time-scale for SN feedback.  This of course also assumes that a single SN will
destroy a GMC instantaneously, while in reality the expanding SN blast wave will
take $>1\Myr$ to reach the edge of a GMC, depending on how much SN energy is
lost due to radiative cooling.

Molecular cloud disruption (feedback) time-scales shorter than 3.5~Myr cannot be
explained by SN feedback.  As \citet{MacLow1988} first showed, SN detonations
from sufficiently large populations (where the number of supernovae $N_{\rm SN}
\gg 1$) can overlap and thermalize, forming a superbubble that evolves as though
it were being driven by a source of constant luminosity.  For typical IMFs
\citep{Kroupa2001,Chabrier2003}, $N_{\rm SN}\sim M_*/(100\Msun)$.  This gives us
two limits that bracket the specific energy injection by SN: an instantaneous
injection of $E_{\rm SN} = 10^{49}(M_*/\Msun)\erg$ at $3.5\Myr$, or a constant
luminosity of $L_{\rm SN} = 1.2\times10^{34}(M_*/\Msun)\ergs$ from
$3.5{-}30\Myr$.  

\subsection{Stellar winds}
Prior to their destruction through core-collapse supernovae, massive stars
inject significant energy into the surrounding natal material through
line-driven stellar winds.  Absorption lines from material in the outer layers
of the stellar atmosphere can make that layer optically thick to those specific
line frequencies \citep{Mokiem2007}.  This in turn couples the photon momentum
of the star's light to these outer layers, driving off a fast stellar wind
($v\sim1000\kms$) which shocks the surrounding gas.  This hot, over-pressured
bubble will then expand into the surrounding ISM as a luminosity-driven blast
wave \citep{Weaver1977}, or as a momentum-driven blast wave if radiative cooling
is efficient \citep{Lancaster2021a,Lancaster2021b}.

As stellar winds are primarily driven by metal absorption lines in their
atmospheres, the total energy input from stellar winds is quite sensitive to
their metallicities.  Both the mass loss rates \citep{Mokiem2007} and the
terminal wind velocities \citep{Leitherer1992} increase with greater
metallicities, giving an approximately constant mechanical luminosity that
depends roughly linearly on the metallicity of the star.  

\subsection{Direct radiation pressure}
Massive stars may also inject momentum into a cloud through radiation pressure
on dust and gas.  Unlike stellar winds, however, there are two different
mechanisms through which radiation pressure can inject energy and destroy a
molecular cloud.  The first is simply through the direct transfer of momentum
through absorbed photons.  In a medium with absorption optical depth $\tau_{\rm
abs}$, a stellar luminosity of $L$ injects momentum at a rate of $\dot
p=[1-\exp{(-\tau_{\rm abs})}]L/c$.  In the limit of an optically thick medium,
this reduces simply to $\dot p = L/c$.  For photon scattering, the amount of
momentum injected by radiation scales proportionally to the scattering optical
depth $\tau_{\rm scat}$, such that $\dot p = \tau_{\rm scat}L/c$.  This means
that for sufficiently large optical depths, the coupling of photon momentum to
the ambient gas  is expected to be dominated by scattering rather than direct
absorption.

In the case of a molecular cloud, ultra-violet (UV) photons harder than the
dissociation energy of molecular hydrogen (Lyman-Werner radiation at $\sim 6
{\rm eV}$) will be absorbed by ${\rm H}_2$ molecules, breaking them apart
\citep{Christensen2012}.  This gives GMCs a very high optical depth to UV
radiation beyond the Lyman-Werner bands, giving $\dot p = L_{\rm UV}/c$ due to
the UV emission of a stellar population.  Dust will also absorb UV photons and
re-radiate the energy imparted by the UV photons in IR bands. IR photons can
then scatter on dust grains within the cloud, which in high IR optical depths
leads to an enhanced momentum injection $\dot p = \tau_{\rm IR}L/c$ as photons
experience multiple scatterings \citep{Krumholz2009}.

\subsection{Photoionization and HII regions}
UV photons emitted by massive stars will not only impart their momentum on the
gas of a cloud, but also photoionize it, breaking apart molecular gas and
raising its temperature to $\sim10^4\K$.  Early in the evolution of this ionized
bubble, ionizing photons will outpace any hydrodynamic expansion of the HII
region, but will rapidly reach a state where the recombination rate at the edge
of the bubble matches the ionization rate set by the stellar UV flux. This
transition begins once an ionization front, driven by an ionizing flux $S$ into
a medium of number density $n_{\rm H}$, with a volumetric case-B recombination
rate $\beta$ reaches the \citet{Stromgren1939} radius  
\begin{equation}
    r_{\rm s} = \left(\frac{3S}{4\pi\beta n_{\rm H}^2}\right)^{1/3} .
\end{equation}
These two phases are described in the classic work of \citet{Kahn1954} as R-type
(ionization driven) and D-type (pressure driven) fronts. As the Str\"omgren
radius in realistic clouds is significantly smaller than the cloud radius,
D-type HII region expansion will be the dominant photoionization feedback for
most of a GMC lifetime (this has been confirmed in 3D radiation hydrodynamic
simulations of HII regions, e.g.\ \citealt{Dale2005,Walch2012,Geen2015}).  If we
are considering momentum generation, R-type HII fronts do not actually generate
significant momentum, as they ionize gas faster than gas pressure can actually
accelerate the cold, molecular gas that surrounds an ionized HII region.  The
dynamics of D-type HII fronts were first derived by \citet{Spitzer1978}.  As the
ionized interior of an HII region is overpressured, it will drive a shock into
the surrounding medium, generating momentum as it sweeps up material into a
shell with radius 
\begin{equation}
    r(t) \propto R_{\rm s}^{3/7}t^{4/7} .
\end{equation}
These results have been extended by \citet{Franco1990} to cover the expansion of
spherically-symmetrical HII regions in clouds with a power-law density profile.
Recent numerical simulations have shown that the non-spherical, turbulent
structure of a GMC may have unpredictable impacts on the leakage of both ionized
gas and ionizing photons from an HII region, and lower the expansion rate of a
D-type HII region.  \citet{Geen2018} examined clouds with identical
globally-averaged properties, but with different (random) IMF sampling and
turbulent driving.  They find that while the average density profiles
surrounding stars formed in GMCs roughly follows a power-law profile, there can
be deviations of $\sim0.5$ dex in the interquartile range of density at
different radii.  This can lead to a factor of $\sim2$ scatter in momentum
injected by HII regions a few $\Myr$ after star formation begins.

\section{Momentum generation by self-similar feedback fronts}
\label{momentum_derivation}

\subsection{General solution}
We can begin by looking for self-similar solutions for the evolution of a
feedback front.  Solutions of this kind will take the form
\begin{equation}
    r\propto t^\alpha .
    \label{selfsim}
\end{equation}
We can use the cloud radius $r_{\rm cl}$ and the feedback time-scale $t_{\rm
FB}$, which is defined as the time needed for the feedback front to reach the
cloud radius \citep{Kruijssen2018}, to write the dimensionless version of this
proportionality as
\begin{equation}
    \frac{r}{r_{\rm cl}} = \left(\frac{t}{t_{\rm FB}}\right)^\alpha .
    \label{selfsim2}
\end{equation}
This simple form can conveniently be used to describe a wide variety of
expanding blast waves, winds, and shells.  A general derivation was presented in
\citet{Ostriker1988} for astrophysical blast waves driven by various driving
sources. As \citep{Ostriker1988} showed, for energy injection by a mechanical
luminosity given by
\begin{equation}
    L_{\rm in} = L_0(t/t_0)^{\alpha_{\rm in}-1} 
\end{equation}
the solution for $\alpha$ in equation~(\ref{selfsim}) will depend only on the
power-law exponent setting the injection rate form $(\alpha_{\rm in})$, with a
functional form depending on the actual driving mechanism.  For most feedback
mechanisms, the specific injection luminosity for a fully-sampled IMF will be
roughly constant \citep{Agertz2013} prior to the first SN explosion, and thus we
will find $\alpha_{\rm in}=1$ if massive star formation within a cloud occurs
$\ll t_{\rm FB}$.  If this star formation continues at a constant rate for the
duration of $t_{\rm FB}$, we will instead have $\alpha_{\rm in}=2$.  For an
adiabatic wind, where radiative losses of the hot bubble (but not necessarily
the swept-up shell) are small, we have $\alpha=(2+\alpha_{\rm in})/5$
\citep{Weaver1977, Ostriker1988}.  If the hot bubble loses most of its thermal
energy to radiation, instead we have $\alpha=(1+\alpha_{\rm in})/4$
\citep{Cioffi1988,Ostriker1988}.  Recent studies of wind cooling in multi-phase,
fractal clouds have suggested that the increased surface area of a fractal
feedback front will amplify radiative cooling losses, but the expansion exponent
$\alpha$ will still follow the overall evolution of an adiabatic or radiative
shell, with the losses captured in a linear coefficient less than unity in the
momentum injection rate \citep{Fielding2020,Lancaster2021a,Lancaster2021b}.
Feedback fronts driven by radiation pressure will follow the evolution of
radiative winds, with the same expansion exponent $\alpha$ as provided above. If
a bubble is not driven by a thermal wind, but instead by the pressure of a
photoionized HII region, \cite{Franco1990} showed that we can use the same
self-similar solution, but this time with $\alpha=(3+\alpha_{\rm in})/7$.  In
table~\ref{momentum_alpha}, we show the different values that the expansion
exponent $\alpha$ can take, for different driving mechanisms with either
instantaneous or constant star formation. As this table shows, the value of
$\alpha$ is constrained to a small range $\alpha=0.5-0.8$.

\begin{table*}
    \begin{tabular}{|l|c|c|}
        Driving Mechanism & Instantaneous SF $(\alpha_{\rm in}=1)$ & Continuous SF $(\alpha_{\rm in}=2)$ \\
        \hline
        \hline
        Adiabatic Wind; $\alpha=(3+\alpha_{\rm in})/5$ & $\alpha=3/5$; $\dot p \propto
        t^{2/5}$ & $\alpha=4/5$; $\dot p \propto t^{6/5}$ \\
        Radiative Wind; $\alpha=(2+\alpha_{\rm in})/4$ & $\alpha=1/2$; $\dot p \propto
        {\rm const}$ & $\alpha=3/4$; $\dot p \propto t$ \\
        D-type HII Shell; $\alpha=(4+\alpha_{\rm in})/7$ & $\alpha=4/7$; $\dot p \propto
        t^{2/7}$ & $\alpha=5/7$; $\dot p \propto t^{6/7}$ \\
        \hline
    \end{tabular}
    \caption{Self-similar expansion exponents $\alpha$ and the momentum
    injection rate for different feedback front driving mechanisms for
    instantaneous star formation $(\alpha_{\rm in} = 1)$ and continuous star
    formation.  As this table shows, the range of values of $\alpha$ varies only
    slightly, from $0.5-0.8$.}
    \label{momentum_alpha}
\end{table*}

As the mass of the feedback front is mostly confined to the swept-up shell
\citep{Weaver1977}, the mass of this shell can be written as a function of the
cloud ambient density $\rho_0$, feedback timescale $t_{\rm FB}$, cloud radius
$r_{\rm cl}$, and cloud-scale star formation efficiency $\epsilon_{\rm SF}$,
given as
\begin{equation}
    M_s(t) = \frac{4\pi\rho_0(1-\epsilon_{\rm SF})r(t)^3}{3} \equiv 
    \frac{4\pi\rho_0 (1-\epsilon_{\rm SF}) r_{ cl}^3}{3} \left(\frac{t}{t_{ FB}}\right)^{3\alpha} .
\end{equation}
the momentum carried by the shell will is then derived simply as the product of
this equation and the time derivative of equation~\ref{selfsim2},
\begin{equation}
    p(t) =M_s(t)v(t) \equiv
    \frac{4r_{\rm cl}\alpha(1-\epsilon_{\rm SF})\pi\rho_0r_{\rm cl}^3}{3t_{\rm FB}}\left(\frac{t}{t_{\rm FB}}\right)^{4\alpha-1} .
\end{equation}
Converting this to a specific momentum per unit stellar mass, where the stellar
population mass is simply $m_*=(4/3)\pi\rho_0\epsilon_{\rm SF}r_{\rm cl}^3$,
gives us
\begin{equation}
    \mathbb{P}(t) =
    \alpha\frac{r_{\rm cl}(1-\epsilon_{\rm SF})}{\epsilon_{\rm SF}t_{\rm FB}}\left(\frac{t}{t_{\rm FB}}\right)^{4\alpha-1}
    \equiv \alpha p_0\left(\frac{t}{t_{\rm FB}}\right)^{4\alpha-1} ,
    \label{momentum_time}
\end{equation}
where the second equality defines $p_0=r_{\rm cl}(1-\epsilon_{\rm
SF})/\epsilon_{\rm SF}t_{\rm FB}$. The momentum injection rate is thus
\begin{equation}
    \mathbb{\dot P}(t) =
    (4\alpha^2-\alpha)\frac{p_0}{t_{\rm FB}}\left(\frac{t}{t_{\rm FB}}\right)^{4\alpha-2} ,
    \label{momentum_rate}
\end{equation}
which is suitable for a direct implementation in hydrodynamical simulations of
galaxy formation and evolution, once the observationally-constrained parameters
$p_0$ and $t_{\rm FB}$ have been provided.

\begin{figure}
    \includegraphics[width=\hsize]{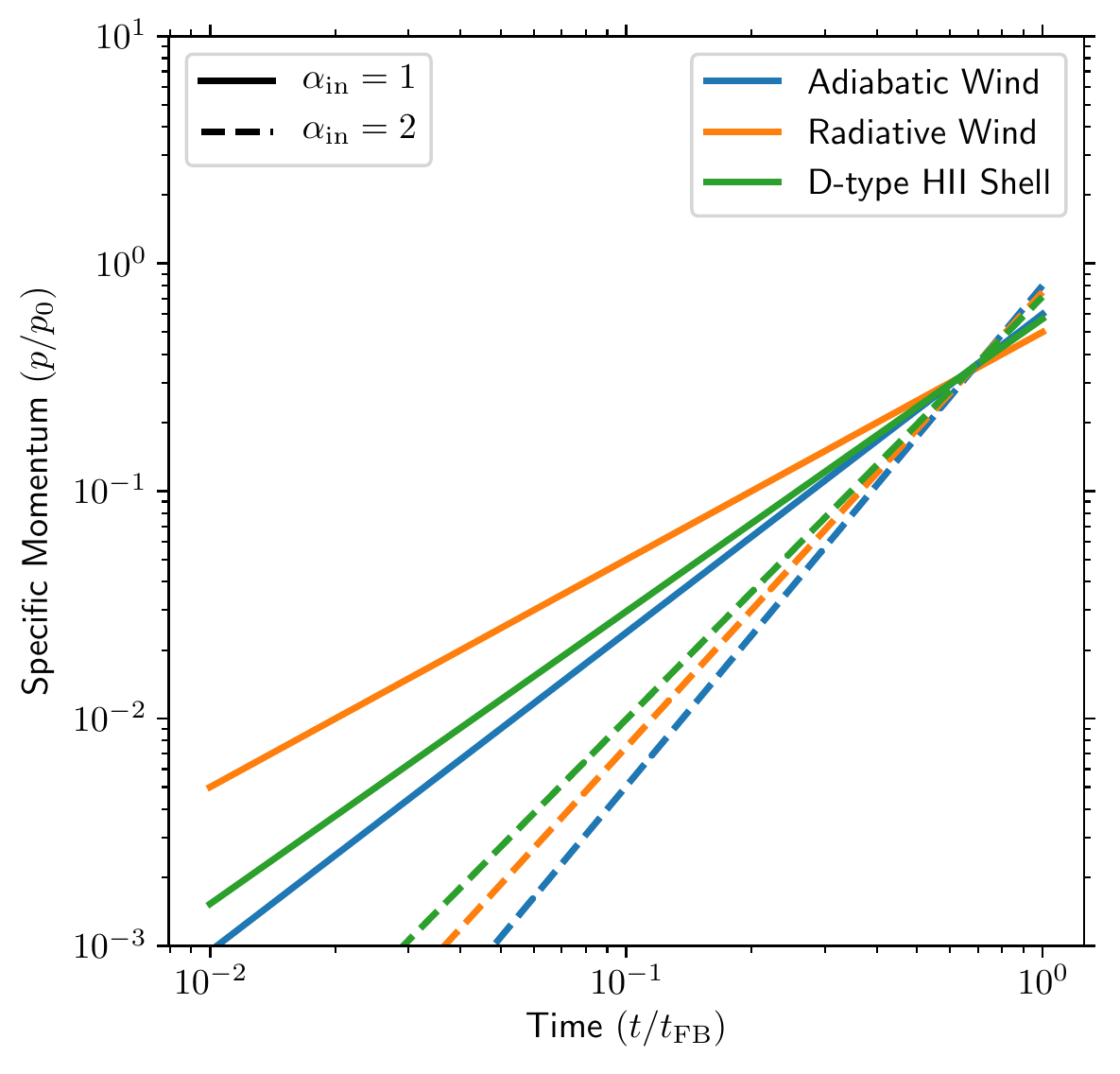}%
    \caption{Momentum injected by different driving mechanisms as a function of
    time.  As can be seen, changing the star formation from instantaneous to
    continuous or the mechanism driving the feedback front changes the slope of
    the momentum injection, but produces less than a factor of 2 difference in
    the final momentum injected.  With instantaneous star formation, we
    naturally see more momentum injected earlier.}
    \label{alpha_rates}
\end{figure}

The momentum injection rates for each of the mechanisms are shown as a function
of time in Figure~\ref{alpha_rates}.  As can be seen, while the early evolution
naturally is more sensitive to the uncertain driving mechanism, the constrained
range of possible expansion exponents results in little variation in the final
momentum injected.

\subsection{Observational measurements of feedback parameters}
\label{obs_measurements}
With the equations derived in the previous section, we can construct the
momentum input rate from observations of cloud radii $r_{\rm cl}$, feedback
time-scales $t_{\rm FB}$, and star formation efficiencies $\epsilon_{\rm SF}$.
To turn this into a two-parameter problem that only requires $p_0$ and $t_{\rm
FB}$, we use the calculation of $p_0=r_{\rm cl}(1-\epsilon_{\rm
SF})/\epsilon_{\rm SF}t_{\rm FB}$ that is part of the {\sc
Heisenberg}\footnote{The code is publicly available at
\url{https://github.com/mustang-project/heisenberg}.} code
\citep{Kruijssen2018}, which employs a Monte-Carlo procedure to
self-consistently propagate the uncertainties on $p_0$.

At the time of writing, the measurements of $p_0$ and $t_{\rm FB}$ with {\sc
Heisenberg} have been made for 15 nearby galaxies
\citep{Kruijssen2019b,Chevance2020,Chevance2022,Ward2020,Zabel2020,Kim2021}, and
will soon be extended to a total sample of $>50$ galaxies (J.~Kim et al.\ in
prep.). For the purpose of this work, we use the homogeneous analysis performed
on CO and H$\alpha$ observations of 10 nearby galaxies, published by
\citet{Kruijssen2019b} and \citet{Chevance2020,Chevance2022}. This sample
includes NGC300, as well as eight galaxies from PHANGS\footnote{PHANGS is
Physics at High Angular Resolution in Nearby GalaxieS, more information is
available at \url{http://phangs.org}}-ALMA (\citealt{Leroy2021}; NGC0628,
NGC3351, NGC3627, NGC4254, NGC4303, NGC4321, NGC4535, NGC5068) and one from
PAWS\footnote{PdBI Arcsecond Whirlpool Survey,
\url{https://www2.mpia-hd.mpg.de/PAWS/PAWS/Home.html}.}
(\citealt{Schinnerer2013}; NGC5194). For each galaxy, the measurements were made
across several bins in galactocentric radius, resulting in a total of 33
independent measurements of $t_{\rm FB}$ and $p_0$ across the 10 galaxies used.
For more details on the analysis, data processing, and radial binning, see
\citet{Kruijssen2019b} and \citet{Chevance2020,Chevance2022}.

\begin{figure}
    \includegraphics[width=\hsize]{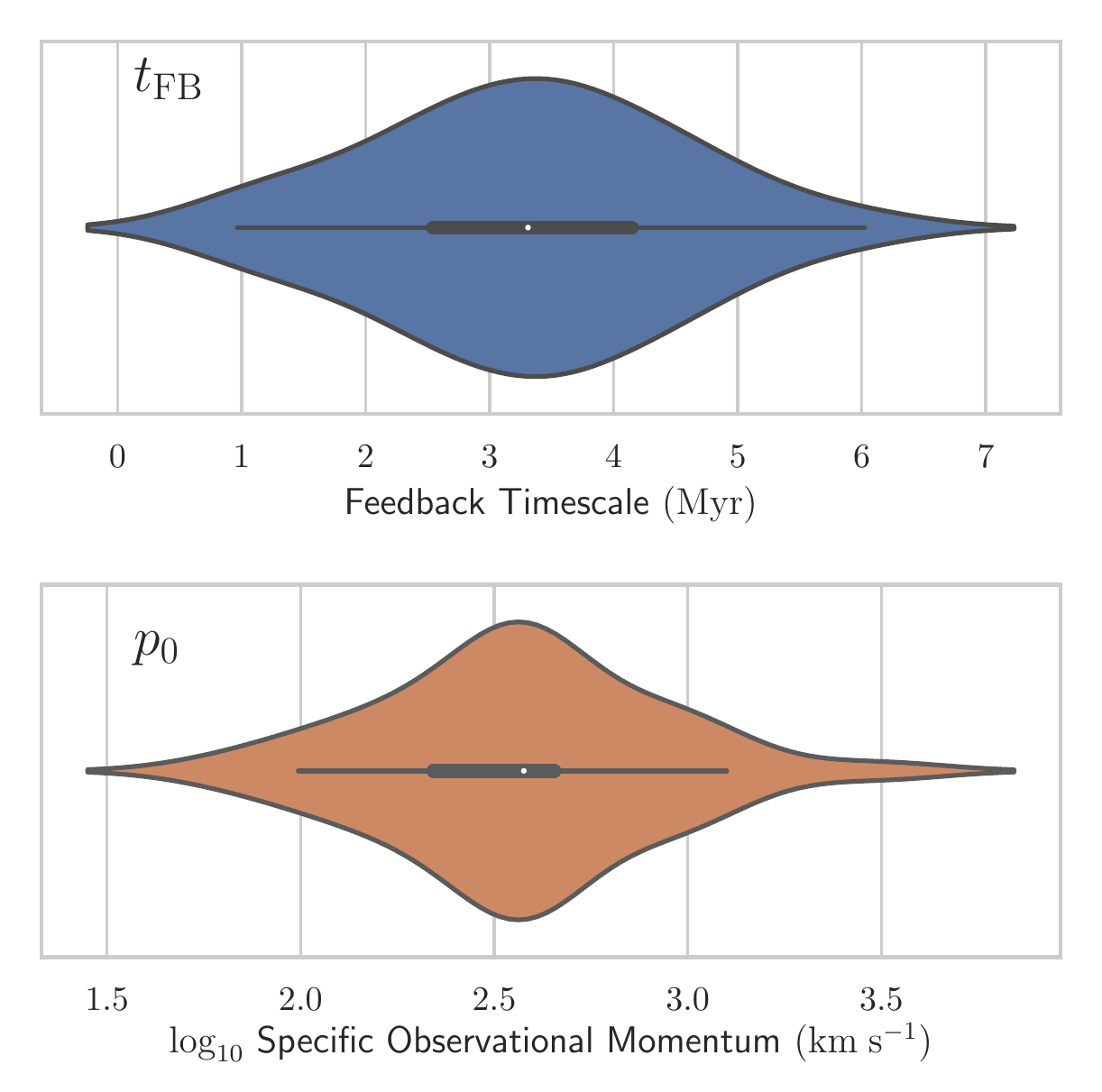}%
    \caption{Violin plots of the two observationally-constrained parameters in
    EMF, $t_{\rm FB}$ (top panel) and $p_0$ (bottom panel).  The coloured
    regions show a kernel density estimate of the observed values.  The thick
    inner lines connect the upper and lower quartiles, while the thin lines show
    the full extents of the measurements.  The white points in the centres show
    the median values ($t_{\rm FB}=3.3\Myr$ and $p_0=377\kms$, respectively).}
    \label{obs_p0}
\end{figure}

It is plausible that $t_{\rm FB}$ and $p_0$ vary with the local properties of
the GMC population and the galactic environment. However, a sample of 33
independent measurements is insufficient to definitively establish such
environmental variations to the required statistical significance. Instead, this
will require the same measurements to be made for the complete PHANGS sample
(J.~Kim et al.\ MNRAS submitted). Therefore, here we do not link the values of $t_{\rm
FB}$ or $p_0$ to either local or galaxy-scale properties, but instead examine
the effect of EMF using the averaged parameters from the current sample of 10
galaxies.  This gives us median values of $t_{\rm FB}=3.31_{-0.76}^{+0.83}\Myr$
and $p_0=377_{-155}^{+74}\kms$ (where uncertainties are the 25$^{\rm th}$ and
75$^{\rm th}$ percentiles).  The distributions of these two parameters are
visualised in Figure~\ref{obs_p0}.

\section{Numerical implementation in galaxy simulations}
\label{simulations}
\subsection{Methods and initial conditions}
We implement a sub-grid model for stellar feedback based on the momentum
injection rates from Section~\ref{momentum_derivation} into the moving-mesh
semi-Lagrangian code {\sc Arepo} \citep{Springel2010}. {\sc Arepo} solves
Euler's equations for hydrodynamics using the Godunov method on a Voronoi mesh
generated on-the-fly using mesh generating points that follow the fluid flow in
each cell.  This method allows for low dissipation and second-order spatial
integration accuracy through the use of a Riemann solver to calculate fluxes
between Voronoi cells, while maintaining Galilean invariance.

{\sc Arepo} has been applied to study galaxy formation in large-volume
cosmological simulations \citep{Vogelsberger2014b}, cosmological zooms of
individual galaxies \citep{Grand2017}, isolated galaxies \citep{Smith2018}, and
stratified slices of the ISM \citep{Simpson2016}.  Radiative cooling in our
version of {\sc Arepo} is handled by the {\sc grackle} 3.1 \citep{grackle}
cooling library, which allows us to include primordial \& metal line cooling
using tabulated {\sc CLOUDY} \citep{Ferland2013} rates. We use {\sc grackle} in
equilibrium mode with an initial ISM metallicity of $Z=0.012$.  The UV
background used in these simulations is derived from \citet{Haardt2012}, at
$z=0$.  We also include a non-thermal pressure floor to ensure that the
\citet{Truelove1997} criterion is not violated, preventing numerical
fragmentation at the resolution scale.  Star formation in our simulations uses a
standard Schmidt-law prescription, with stars forming from gas cells at a rate
set by $\dot \rho_* = \epsilon_{\rm ff}\rho/t_{\rm ff}$, where $t_{\rm ff}$ is
the gas free-fall time, and $\epsilon_{\rm ff}$ is the dimensionless star
formation efficiency (per free-fall time) parameter.  We allow gas cells with density above $100\hcc$
and temperature below $10^4\K$ form stars, with a local star formation
efficiency per free-fall time of $\epsilon_{\rm ff} =0.1$.  While this value is
near the upper range of observed cloud-scale star formation efficiencies
\citep{Evans2009,Heyer2016,Grudic2019}, it should be noted that the efficiency
of the sub-grid star formation model $\epsilon_{\rm ff}$ does not account for
the effect of feedback (which we explicitly model in our simulations).  The
actual cloud-scale star formation efficiency in these simulations are therefore an emergent
property of gravity, hydrodynamics, and feedback in concert
\citep[e.g.][]{Grisdale2019}.

We simulate an isolated, Milky-Way like galaxy with initial conditions (ICs)
drawn from the AGORA comparison project \citep{Kim2014}.  The AGORA isolated
disc IC has a disc scale radius of $3.43\kpc$ and a scale height of $343\pc$.
The disc is embedded in a dark matter (DM) halo with a mass of
$M_{200}=1.07\times10^{12}\Msun$ and a virial radius of $R_{200}=205\kpc$.  The
halo concentration parameter is $c=10$, with a \citet{Bullock2001} spin
parameter of $\lambda=0.04$.  The galaxy model contains both a stellar disc and
bulge, with a bulge-to-disc ratio of $0.125$ and a total gas fraction of $0.18$.
The AGORA disc ICs were generated using the {\sc MakeNewDisk} code
\citep{Springel2005}.  We use a gravitational softening length of $40\pc$ and
$260\pc$ for baryons and DM respectively, and a gas cell mass of
$8.59\times10^3\Msun$.   In Appendix~\ref{convergence}, we show that the star
formation and feedback quantities we measure here are well-converged at this
resolution. The IC star particle mass is $3.437\times10^4\Msun$, and the live DM
halo contains $10^5$ particles of mass $1.254\times10^6\Msun$ each.  We use
Lagrangian refinement to ensure that individual cell masses never deviate by
more than a factor of two from the target mass resolution. We initialize the gas
in the simulation with a temperature of $10^4\K$, though this is rapidly
replaced with the equilibrium temperature calculated by {\sc grackle}.

\subsection{Numerical implementation of empirically-motivated early feedback}
We implement EMF as a model with three parameters.  The first two are $t_{\rm
FB}$ and $p_0 = r_{\rm cl}(1-\epsilon_{\rm SF})\epsilon_{\rm SF}^{-1}t_{\rm
FB}^{-1}$ and are constrained empirically, while the third parameter is
$\alpha$, which needs to be chosen from a range of possible values (see
table~\ref{momentum_alpha}). Following equation~(\ref{momentum_time}), during a
timestep $\Delta t$ star particles with birth mass $M_{\rm i}$ and age $t<t_{\rm
FB}$ will inject an amount of momentum
\begin{equation}
    \Delta p(M_{\rm i}, t, \Delta t) = \alpha p_0 M_{\rm i} \left[\left(\frac{t+\Delta
    t}{t_{\rm FB}}\right)^{4\alpha-1}-\left(\frac{t}{t_{\rm
    FB}}\right)^{4\alpha-1}\right] .
\end{equation}
Naturally, this is a piecewise linear approximation of the momentum injection
rate of equation~(\ref{momentum_rate}).  The momentum generated by each star
particle is deposited into the surrounding cells by first finding the cell in
which a star particle resides, and using the Voronoi mesh to weight the
contribution to each of the cells by their relative area. Any momentum
cancellation that occurs due to existing motions of the surrounding gas cells is
thermalized, and this thermal energy is deposited in the star particle's host
cell.  The cell that a star particle resides in has a total surface area $A_i$,
and its cell faces shared with each surrounding cell contribute an area $S_j$,
such that $A_i = \Sigma_j S_{ij}$, giving a weight of $W_{ij}=S_{ij}/A_i$ for
each neighbouring cell.  Thus, each star particle contributes to a cell $j$
momentum $p_j = p(M_{\rm i}, t, \Delta t)W_{ij}$, in the direction of the normal
vector to the surface $S_{ij}$.  This guarantees that the total {\it absolute}
momentum injected is $p(M_{\rm i}, t, \Delta t)$, while the total vector
momentum is zero by the divergence theorem (thus ensuring momentum
conservation).  This algorithm can be trivially adapted for both fixed-mesh
Eulerian codes and Smoothed-Particle Hydrodynamics (SPH) Lagrangian codes,
simply by replacing the weighting factor $w_j = S_j/A$ with the Cartesian
equivalent or an SPH kernel weight.  

As discussed in Section~\ref{obs_measurements}, the parameters $p_0$ and $t_{\rm
FB}$ are derived from the observational data ($p_0=377\kms$, $t_{\rm
FB}=3.3\Myr$).  As the exponent $\alpha$ is unconstrained observationally,
arising from the density structure of GMCs and the feedback driving mechanism,
we simulate cases with two different values of $\alpha=\{0.5, 1.0\}$.  This
range bounds all reasonable possible values of $\alpha$, as shown in
Table~\ref{momentum_alpha} and and discussed in
Section~\ref{momentum_derivation}.

Supernova feedback is handled by the mechanical supernovae scheme first
described in \citet{Kimm2014}, using the same Voronoi face weighting scheme as
the EMF.  In brief, we calculate a terminal momentum, when a single SN
transitions to the momentum-conserving phase
\begin{equation}
    \frac{p_{\rm term}}{10^5\kms\Msun} = 3E_{51}^{16/17}n_H^{-2/17}{\rm max}(Z/{\rm Z}_\odot,
    0.01)^{-0.14} .
    \label{terminal_mom}
\end{equation}
In this equation, $E_{51}$ is the total energy of all SN detonating within a
cell (in units of $10^{51}\erg$), while $n_H$ is the ambient density of each
cell momentum is injected into.  The mechanical feedback algorithm then
automatically switches between a thermal-dominated and kinetic-dominated
algorithm based on the local resolution.  Of the total SN energy $E_{\rm SN}$
injected into cell $i$, we deposit kinetic energy into neighbouring cells $j$ in
the form of momentum, calculated as
\begin{equation}
    p_{\rm FB} = {\rm min}(\sqrt{2W_{ij}M_jE_{\rm SN}}, W_{ij}p_{\rm term}) .
\end{equation}
Each cell always deposits a total energy of $E_{\rm SN}$, with the kinetic
fraction set by $p_{\rm FB}$.  Kinetic energy is injected into the neighbouring
cells, while the thermal fraction is deposited into the central cell. Thus, if
$M_j$ is large (in the limit of low resolution), we inject the terminal
momentum, leading to a small thermal contribution, while if $M_j$ is small (in
the limit of high resolution), we inject a larger fraction as thermal energy,
which then will generate the terminal momentum self-consistently through gas
pressure acting on the surrounding material.  This method is similar to
\citet{Hopkins2018a} and \citet{Smith2018}, and has been used by
\citet{Jeffreson2020,Jeffreson2021a,Jeffreson2021b} and \citet{Keller2022a}.  We
deposit $10^{51}\erg$ per SN and calculate the chemical enrichment using a
physically-motivated SN distribution generated by {\sc
SLUG}\footnote{Stochastically Lighting Up Galaxies,
\url{https://slug2.readthedocs.io/en/latest/}.} \citep{Krumholz2015}, using rate
tables drawn from \citet{Sukhbold2016}.  In addition to core-collapse SNII, we
also deposit mass, metals, and energy from SNIa using a delay-time distribution
derived by \citet{Maoz2012} with metal yields from \citet{Seitenzahl2013}.

\begin{figure}
    \includegraphics[width=\hsize]{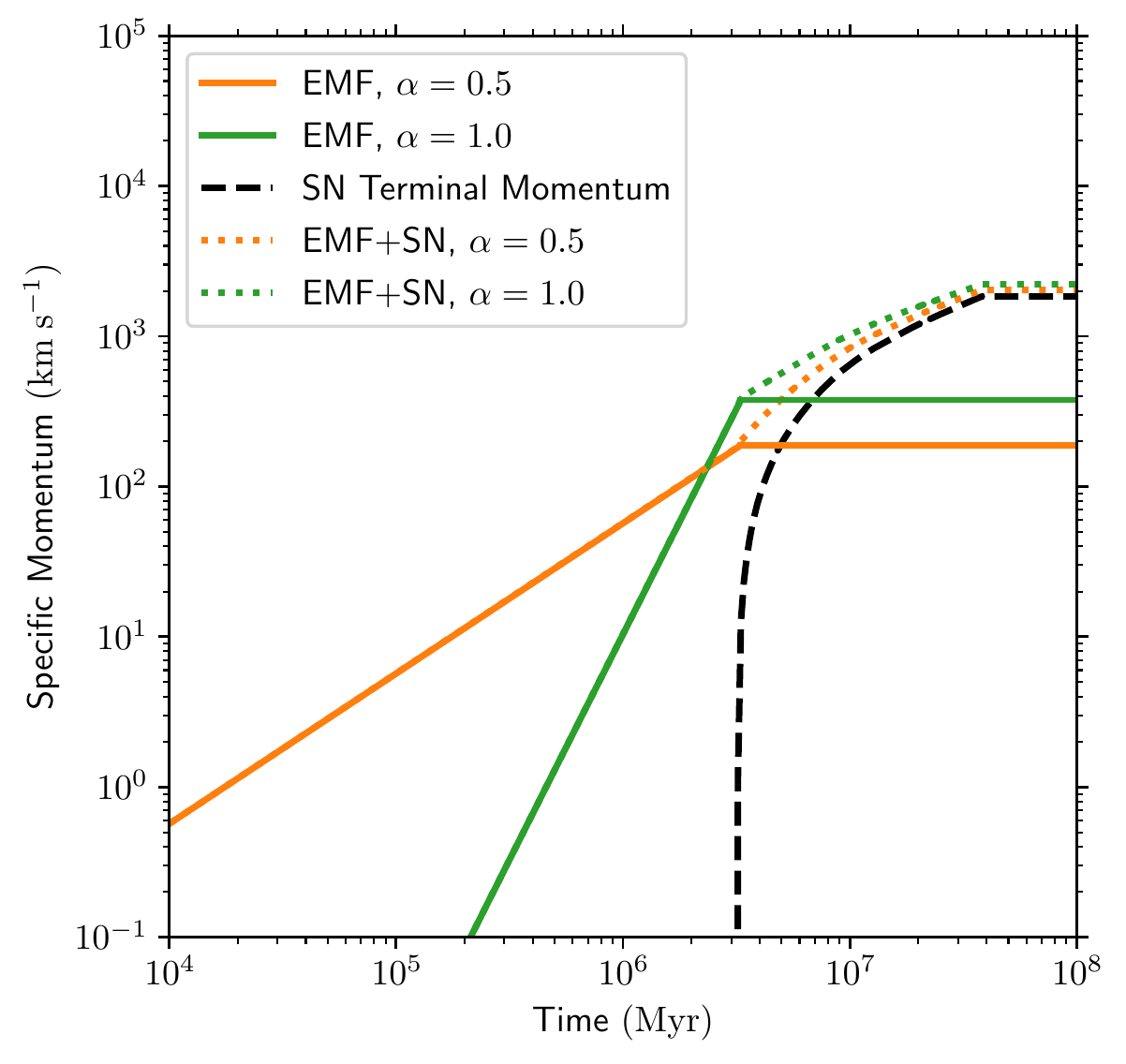}%
    \caption{Total specific momentum injected by SN (black curve) and EMF with
    $\alpha=0.5$ (orange curve) and $\alpha=1.0$ (green curve).  The SN momentum
    is calculated using equation~\ref{terminal_mom}, along with SN rates
    calculated by {\sc Starburst99} for a fully-sampled \citet{Kroupa2001} IMF,
    in an ambient medium with $n_H=1\hcc$ and solar metallicity.  The coloured
    dashed curves show the momentum injected by both EMF and SN together.}
    \label{total_mom}
\end{figure}

In Figure~\ref{total_mom}, we show the specific momentum injected from a stellar
population by both EMF and SN.  As can be seen, the total momentum injected by
EMF with $\alpha=1.0$ is roughly twice the total momentum injected with
$\alpha=0.5$, though most of this momentum is injected later, only exceeding the
momentum injected with $\alpha=0.5$ after $2\Myr$.  The momentum injected by SN
is calculated using equation~\ref{terminal_mom} in an ambient ISM with
$n_H=1\hcc$ and solar metallicity.  While this likely overestimates the momentum
budget of SN at early times, it represents the final momentum that will be
injected at the end of the pressure-driven snowplow phase
\citep{Cioffi1988,Blondin1998}.  As can be seen in Figure~\ref{total_mom}, the
terminal SN momentum is 5-10 times larger than the momentum injected by EMF, but
this momentum only begins being deposited after $\sim3\Myr$.  EMF modifies the
timing of momentum injection, without significantly changing the total momentum
injected by all forms of feedback after the final SN detonates.

\begin{figure}
    \includegraphics[width=\hsize]{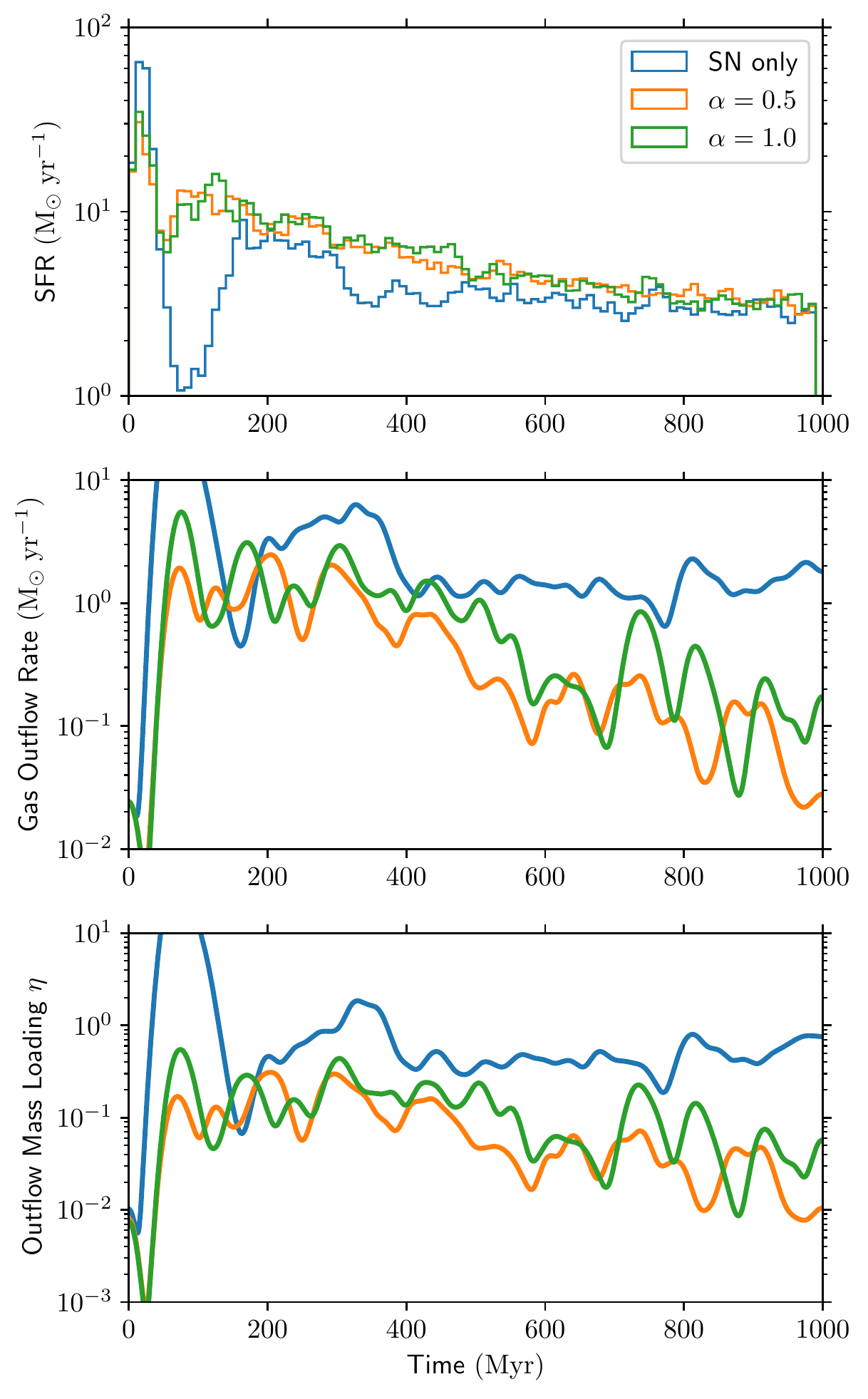}%
    \caption{Star formation rate (top panel), outflow rate (middle panel), and
    mass-loading (bottom panel) for three different isolated disc galaxies (one
    with SN feedback only and two including EMF with $\alpha=\{0.5, 1.0\}$).  As
    the top panel shows, after an initial burst of star formation and settling
    of the disc, the final $500\Myr$ of evolution for all three examples show
    roughly equivalent star formation rates.  However, the outflow rates and
    mass loadings are approximately an order of magnitude lower with the
    addition of EMF.}
    \label{alpha_sfr}
\end{figure}
\begin{figure}
    \includegraphics[width=\hsize]{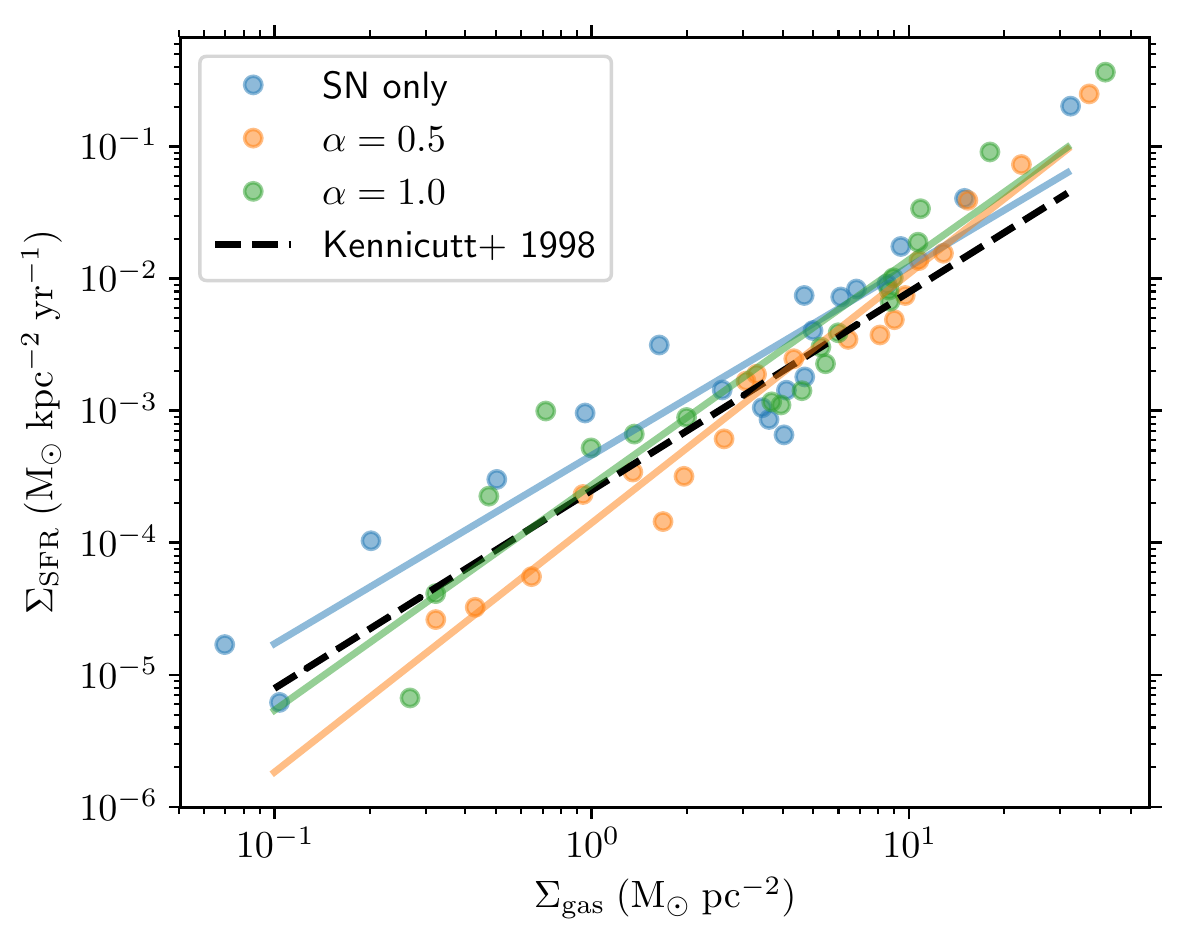}%
    \caption{Kennicutt-Schmidt diagram showing the relation between total gas
    surface density $\Sigma_{\rm gas}$ and star formation surface density
    $\Sigma_{\rm SFR}$ in our simulated galaxies with SN alone (blue points) and
    EMF with $\alpha=0.5$ (orange points) and $\alpha=1.0$ (green points).
    Solid curves show power-law fits to each set of data points.  The dashed
    line shows the \citet{Kennicutt1998} observational relation.  We see
    a slightly lower $\Sigma_{\rm SFR}$ for the galaxies which include EMF when
    $\Sigma_{\rm gas}<1\Msun \pc^{-2}$, and steeper slopes to the fitted
    Kennicutt-Schmidt relations for the EMF galaxies.}
    \label{KS}
\end{figure}
\begin{figure}
    \includegraphics[width=\hsize]{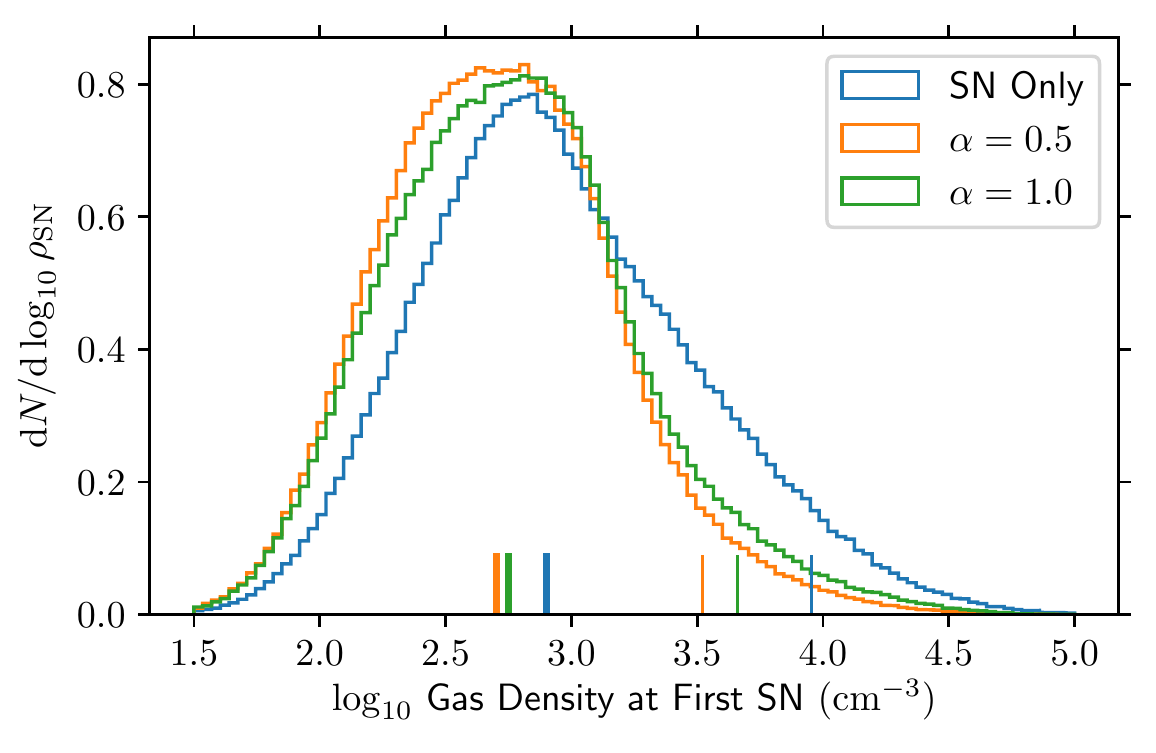}%
    \caption{Probability distribution function of the ambient gas density at the
    location of the first SN event of each star particle, for three simulated
    galaxies (one with SN feedback only and two using EMF with $\alpha=\{0.5,
    1.0\}$).  The thick vertical markers show the median gas densities at the
    first SN event, and the thin vertical markers show the 95$^{\rm th}$
    percentiles, probing the SNe that detonate in the densest environments
    (where they will suffer most from radiative cooling).  As is clear, EMF
    reduces the typical ISM density that SNe detonate within, and in particular
    it shifts the high-density tail of the SN environmental density, lowering
    the 95$^{\rm th}$ percentile by a factor of $2{-}3$.}
    \label{alpha_sn_density}
\end{figure}
\begin{figure}
    \includegraphics[width=\hsize]{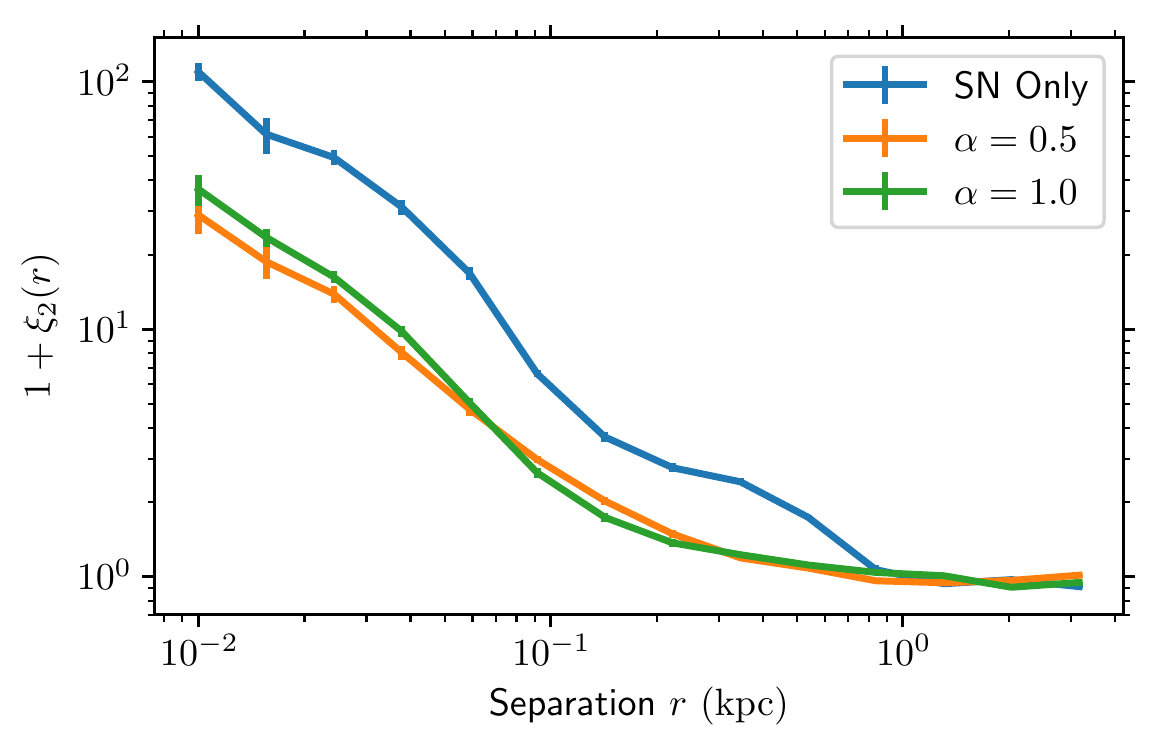}%
    \caption{Two-point correlation function of young ($<20\Myr$) stars, for
    three simulated galaxies (one with SN feedback only and two including EMF
    with $\alpha=\{0.5, 1.0\}$).  While there is no statistically significant
    difference between the small-scale $(<100\pc)$ correlation of young stars
    for different values of $\alpha$, the clustering of stars in the run without
    early feedback is much larger on scales below $1\kpc$.}
    \label{alpha_clustering}
\end{figure}
\begin{figure*}
    \includegraphics[width=\textwidth]{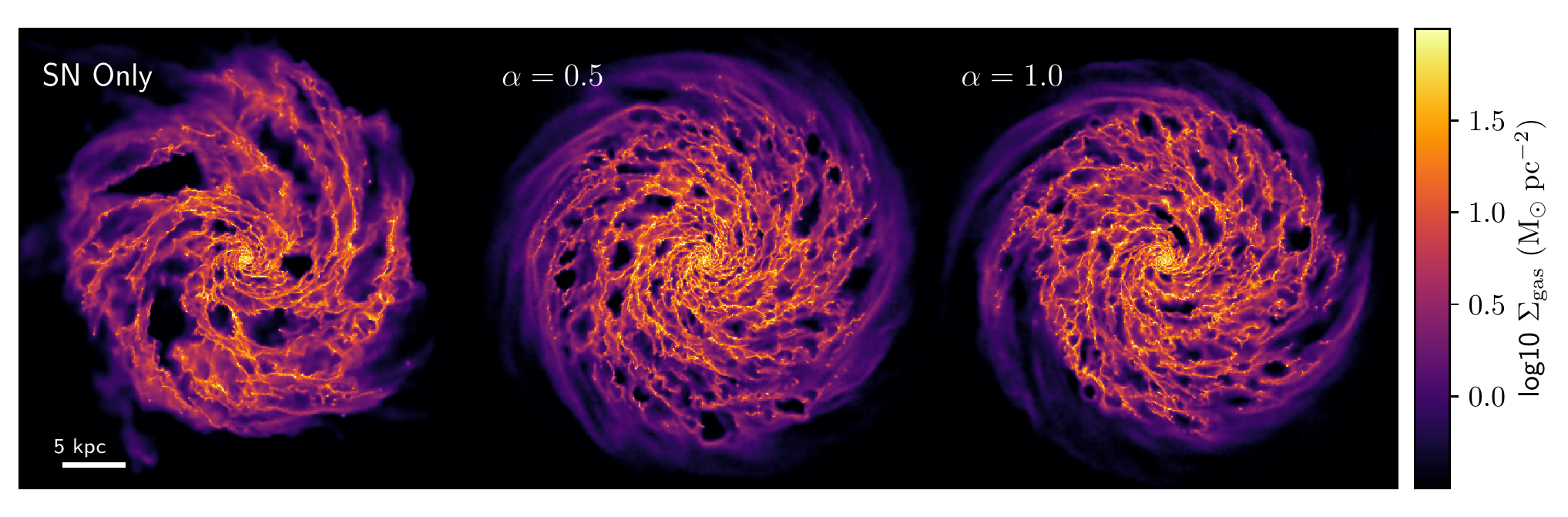}
    \includegraphics[width=\textwidth]{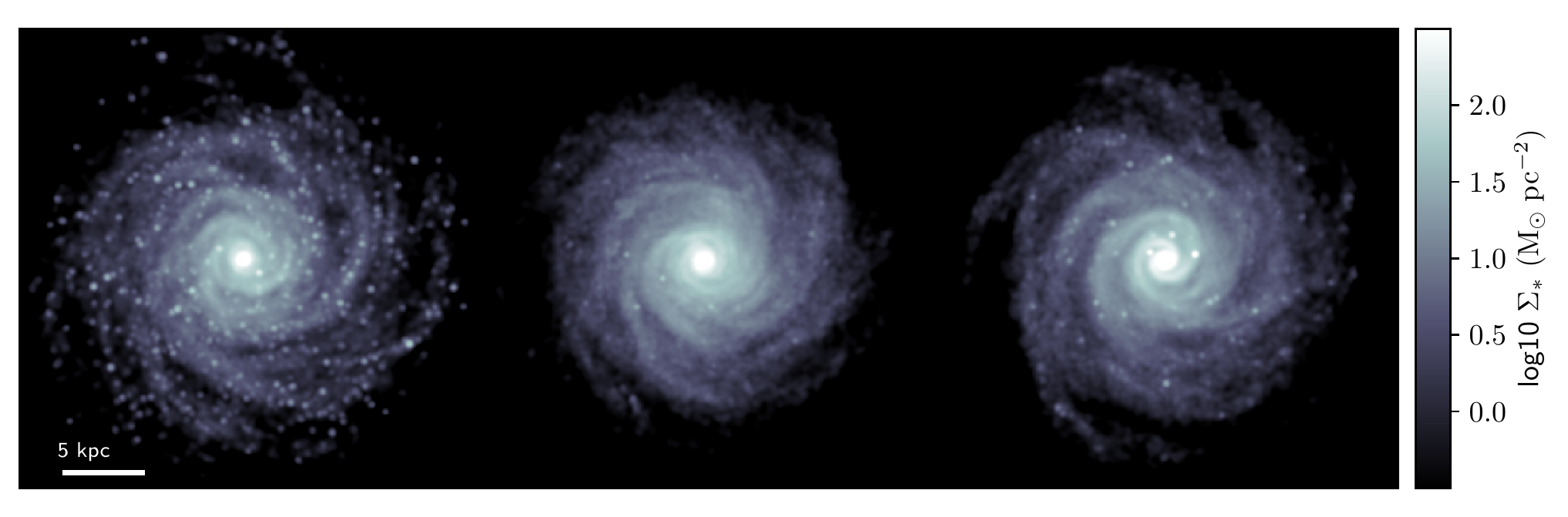}%
    \caption{Gas (top row) and stellar (bottom row) column density maps at
    $t=1\Gyr$ for an isolated disc galaxy with SN feedback only (leftmost
    panel), and when including EMF with $\alpha=\{0.5, 1.0\}$ (see the
    annotations). The ISM is more homogeneous and flocculent when EMF is
    included.  Feedback-driven voids in the ISM are generally smaller, and the
    spiral arm structure is less pronounced.  In the stellar column density, the
    EMF simulations clearly show reduced small-scale clustering, with only a
    handful of dense stellar groups visible in each of the EMF cases. The top
    panels have a width of $40\kpc$, while the bottom panels have a width of
    $30\kpc$.}
    \label{alpha_images}
\end{figure*}

\begin{figure}
    \includegraphics[width=\hsize]{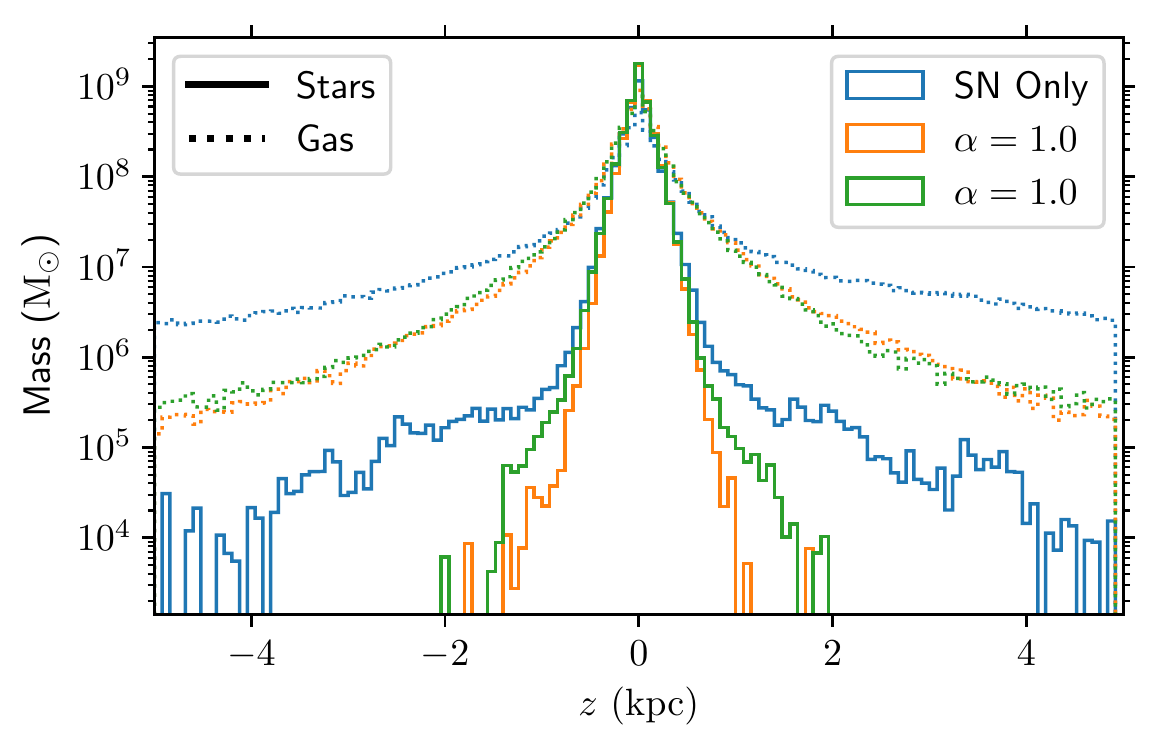}%
    \caption{Vertical profiles of gas (dotted curves) and stars (solid curves)
    for each of our simulated galaxies.  As before, we see little difference
    between different values of $\alpha$ when EMF is included, but there is a
    notable decrease in gas at high latitudes relative to the SN-only
    simulation. This is to be expected from the reduced outflow rates, because
    the isolated nature of these discs implies that gas significantly above the
    initial scale height of $343\pc$ is deposited there primarily through
    outflows.  Interestingly, we also see a significant, highly extended thick
    stellar disc component in simulations without EMF.  With EMF included, no
    stars are found more than $\sim2\kpc$ above or below the disc.}
    \label{alpha_scaleheight}
\end{figure}
\begin{figure}
    \includegraphics[width=\hsize]{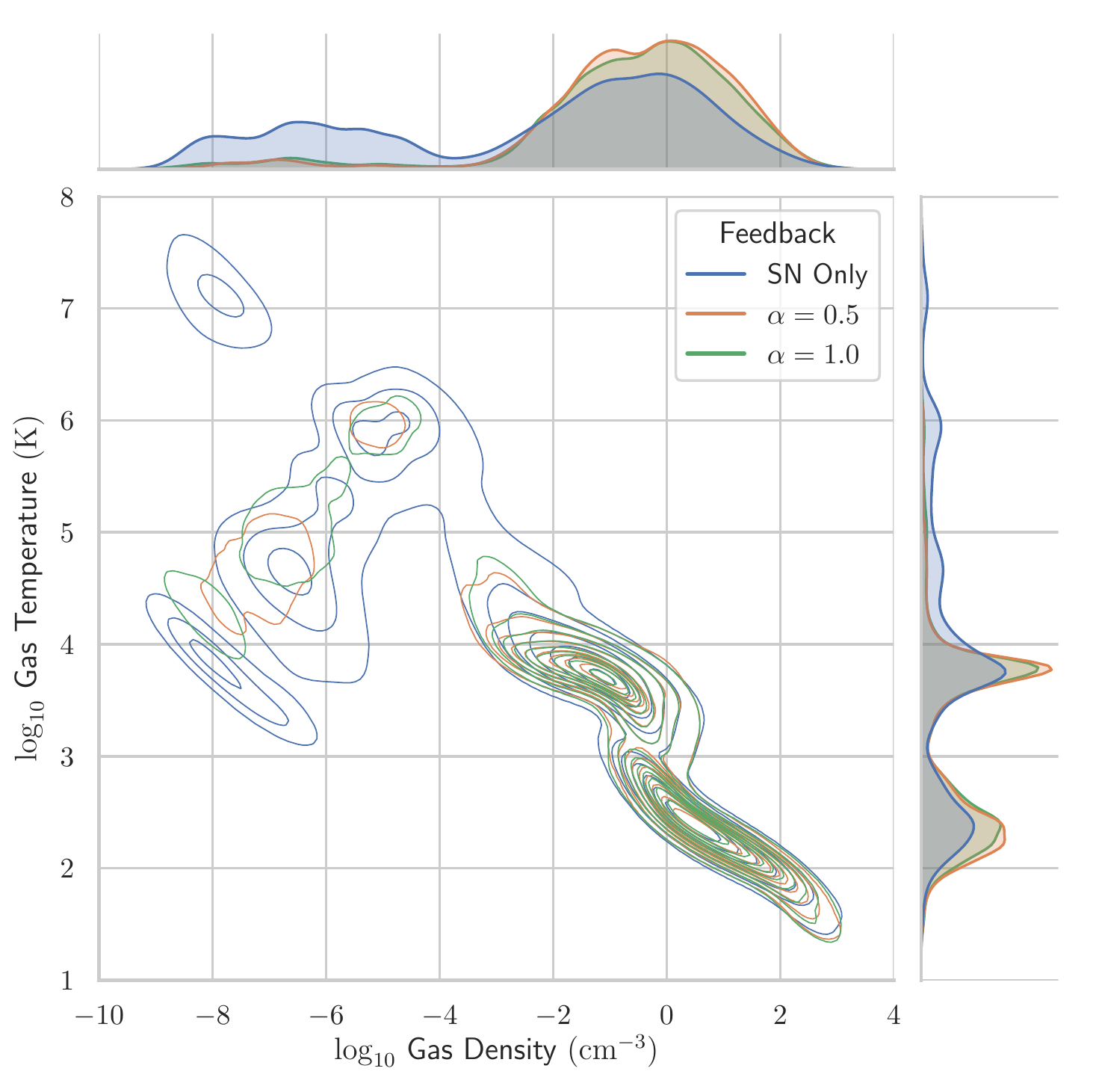}%
    \caption{Density-temperature phase diagram for three simulated galaxies (one
    with SN feedback only and two including EMF with $\alpha=\{0.5, 1.0\}$).
    Each panel shows a kernel density estimate (KDE) weighted by gas mass, for
    each simulated galaxy.  The central panel shows contours of this KDE in the
    density-temperature plane, while the marginal panels each show the
    probability distribution function in density (top panel) and temperature
    (right panel). Each contour set contains ten equally spaced quantiles from
    0.05 to 1 in the KDE probability. As can be seen, there is little difference
    in the distribution of gas in either density or temperature as a function of
    $\alpha$. However, when only including SN feedback, we see significantly
    more gas at high ($>10^5$~K) temperatures and low densities ($<10^3\hcc$).
    This is a result of the reduced radiative losses that occur due to the
    clustering of SN feedback demonstrated in Figure~\ref{alpha_clustering}.}
    \label{alpha_phase}
\end{figure}

\section{Impact of empirically-motivated feedback on galactic structure and the multi-scale baryon cycle}
\label{results}
In this section, we discuss how our new empirically-motivated feedback model
influences the baryon budget of galaxies through the balance between star
formation and outflows, how it affects the structure and properties of the
galactic disc, and how it impacts the molecular cloud lifecycle. Taken together,
this describes how the multi-scale baryon cycle changes when adding EMF.

\subsection{Self-Regulation and Outflows}
The first question we examine is how the addition of EMF changes the overall
regulation of star formation, which is often tied together with the ability of
star formation to drive outflows from the galactic disc
\citep{Keller2015,Rosdahl2017}.  In Figure~\ref{alpha_sfr}, we examine the star
formation rate $\dot M_*$, the gas outflow rate $\dot M_{\rm out}$, and the mass
loading ($\eta=\dot M_{\rm out} / \dot M_*$) for a galaxy simulated with SN
feedback alone, versus two cases of EMF with $\alpha=\{0.5,1.0\}$.  Outflow
rates are calculated in the direction perpendicular to the disc plane, through
two $500\pc$ thick slabs located $5\kpc$ above and below the disc mid-plane.  We
also smooth the outflow rates over a $10\Myr$ window to remove the
high-frequency stochasticity in the outflow rate.  

After a $\sim 400\Myr$ settling period, during which the disc cools and comes
into feedback self-regulation, the star formation rates (SFRs) of the simulated
galaxies are comparable, with a relatively constant SFR of $\sim
3\Msun\yr^{-1}$ in the SN only case, and a slowly declining SFR from
$\sim6\Msun\yr^{-1}$ at 400 Myr to $\sim3\Msun\yr^{-1}$ at $1\Gyr$ in the galaxies
simulated with EMF.\footnote{Interestingly, the reduced initial burst of star
formation when early feedback is included has also been seen in extremely high
resolution simulations of dwarf galaxies by \citet{Smith2021}.  Despite only
lasting $4\Myr$, the initial delay between the first stars forming and SN
feedback occurring appears to be enough to push the disc out of equilibrium in
isolated galaxy simulations, independently of the mass scale and resolution.}
As we might expect, EMF reduces the magnitude of the initial starburst, which is
triggered by the disc cooling out of equilibrium before SN feedback regulation
can moderate the SFR.  Without any form of early feedback, the SN-only run is
able to form stars in newly collapsed GMCs for $\sim4\Myr$ prior to any feedback
energy being injected, which greatly amplifies the initial burst. After $1\Gyr$,
the final stellar mass of the galaxy is $3.32\times10^9\Msun$ in the SN-only
case, versus $\{4.03, 4.21\}\times10^9\Msun$ for $\alpha=\{0.5, 1.0\}$.  EMF
modestly increases the averaged star formation rate by $\sim20\%$ over
simulations with SN alone, which is in part caused by the different response to
the initial starburst. For the final $200\Myr$, long after the effects of the
initial starburst have receded, the average SFR for the SN-only case is
$2.89\Msun\yr^{-1}$, versus $\{3.33, 3.24\}\Msun\yr^{-1}$ for $\alpha=\{0.5,
1.0\}$, a difference of only $11-14$\%.

In contrast with the SFRs and stellar masses, the outflow rates exhibit two
major quantitative differences. First, EMF produces a significantly reduced
average outflow rate and mass loading.  For SN feedback alone, the median
outflow rate for the final $500\Myr$ of the simulation is $1.4\Msun\yr^{-1}$,
while EMF drives median outflow rates of $\{0.11, 0.20\}\Msun\yr^{-1}$ for
$\alpha=\{0.5, 1.0\}$.  Not only is the averaged outflow rate (and mass loading)
reduced by roughly a factor of 10 when EMF is enabled, these outflows also
become more transient, with variations of $\sim 1$ dex over $\sim50-100\Myr$
time-scales.  This may indicate a greater sensitivity to the local ISM
environment in the simulations which include EMF.  Past studies
\citet{Rosdahl2017,Keller2022a} have shown that the detailed modelling of
stellar feedback produces a more significant impact on outflow rates than the
SFR, though the magnitude of the effects depends on the feedback process and
numerical model being considered.

In the Kennicutt-Schmidt diagram shown in Figure~\ref{KS}, we calculate gas and
SFR surface density at $t=1\Gyr$ in 20 radial annuli of $1\kpc$ width between
$0-20\kpc$, excluding gas which is more than $1.5\kpc$ above the galaxy
midplane.  The star formation rates here are calculated using stars younger than
$50\Myr$. We can see that the three cases (SN alone, EMF with $\alpha=0.5$, and
EMF with $\alpha=1.0$), the general trend follows the \citet{Kennicutt1998}
relation for $\Sigma_{\rm gas}<10\Msun \pc^{-2}$, but increases in slope at
higher gas surface densities.  We see a somewhat higher $\Sigma_{\rm SFR}$ at
low gas surface density ($\Sigma_{\rm gas}<1\Msun\pc^{-2}$) when EMF is omitted.
The best-fit Kennicutt-Schmidt relations ($\Sigma_{\rm SFR}=A(\Sigma_{\rm
gas}/\Msun\pc^{-2})^N\Msun\kpc^{-2}\yr^{-1}$) for each case is
$A=4.6\times10^{-4}$, $N=1.4$ for SN alone; $A=1.4\times10^{-4}$, $N=1.9$ for
EMF with $\alpha=0.5$, and $A=2.7\times10^{-4}$, $N=1.7$ for EMF with
$\alpha=1.0$.  With a different number of annuli or pixel-based
Kennicutt-Schmidt surface densities, we see the same general trend: higher SFR
at low surface densities for SN alone, and a steeper best-fit Kennicutt-Schmidt
relation with EMF included.

It may seem puzzling at first that including additional feedback momentum from
pre-SN feedback would reduce the overall effectiveness of stellar feedback to
drive mass-loaded galactic winds and fountains.  To understand this, we need to
examine the environment in which the SNe detonate.  Early feedback can change
the efficiency of star formation regulation through three channels: adding to
the overall energy and momentum budget, changing the ambient density in which SN
detonate (thereby changing cooling losses), and changing the spatial and
temporal clustering of SNe.  In Figure~\ref{alpha_sn_density}, we show how the
ambient gas density around the first SN event changes with the addition of EMF.
We measure the ambient density as the density of the gas cell in which a star
particle finds itself at the time of the first SN event.  By measuring the
ambient density of the first SN event, rather than for all SNe that are produced
by a star particle, we isolate how EMF shapes the initial gas environment that
SNe may detonate in, because the first SN events may change the detonation
environment of later SNe.  As can be seen from the histogram in
Figure~\ref{alpha_sn_density}, EMF causes a slight reduction of the median
density that SNe detonate in. For SNe alone, the median gas density at the site
of the first SN event is $8.0\times10^2\hcc$, compared to $\{5.0,
5.7\}\times10^2\hcc$ when including EMF with $\alpha=\{0.5, 1.0\}$.  The
high-density tail, where SN will experience the most extreme cooling losses,
shows a somewhat stronger trend.  Without EMF, the 95$^{\rm th}$-percentile
ambient density is $9.0\times10^3\hcc$, versus $\{3.3, 4.6\}\times10^3\hcc$ when
including EMF with $\alpha=\{0.5, 1.0\}$. These results alone should point to
{\it greater} efficiency for stellar feedback to drive outflows and regulate
star formation, as lower ambient densities at the site of SN detonation should
yield less radiative losses, and allow more energy to drive gas heating and
acceleration.

Figure~\ref{alpha_sn_density} shows that the reduced outflow rates cannot be
explained by a change in the overall cooling losses of the first SN that might
have been driven by early feedback expelling gas from the SN environment.
However, it has also been shown that clustered SN are much more efficient at
injecting momentum \citep{Gentry2020} and driving galactic outflows
\citep{Fielding2018,Martizzi2020}.  In Figure~\ref{alpha_clustering}, we show
the two-point correlation function, calculated with the
\citet{Landy1993}\footnote{The \citet{Landy1993} estimator uses the number of
true pairs within a separation $r$, $DD(r)$, together with the number of random
pairs with the same mean density $RR(r)$ and the cross-correlated data-random
pairs $DR(r)$ to calculate $\xi_2(r) = (DD(r)-2DR(r)+RR(r))/RR(r)$.  The
\citet{Landy1993} estimator is designed to minimize errors occurring from a
non-periodic distribution of points.} estimator, of young (with ages $<20\Myr$)
stars in each of our simulations.  As can be seen, on length scales below
$800\pc$, the galaxy simulated with SN feedback alone exhibit a significantly
enhanced clustering of young star particles.  The probability of finding pairs
of star particles with separations of $\sim 10\pc$ is up to 4 times greater when
EMF is omitted. This difference follows naturally from the delay of stellar
feedback in the SN-only simulation. SN feedback begins $4\Myr$ after the birth
of a star particle, which corresponds to the typical main sequence lifetime of
the most massive ($>40\Msun$) stars.  This allows additional star formation in
the neighbourhood of young star particles to continue unopposed by the injection
of feedback from star particles that have already formed, but that have not yet
detonated their first SN.  By contrast, EMF regulates star formation on the
cloud scale during the $\sim3\Myr$ after the first star particle forms, thereby
reducing the overall clustering of young stars and diminishing the effectiveness
of SNe at driving galactic outflows.  The same effect has been seen in the
recent, high-resolution dwarf galaxy simulations of \citet{Smith2021}. Despite a
ten-fold reduction in galactic outflows, this local (rather than global)
regulation of star formation produces an averaged galactic SFR that is nearly
indistinguishable from the SN-only case.

\subsection{How EMF reshapes the stellar and gaseous discs}

So far, we have demonstrated changes in the mode of star formation regulation by
the addition of early feedback.  Without early feedback, star formation shows
stronger spatial correlation, which in turn produces larger SN-powered
superbubbles, driving higher outflow rates and producing larger voids in the
ISM.  This changing mode of regulation produces both quantitative and
qualitative differences in the structure of the ISM and the stellar discs of our
simulated galaxies.  The qualitative changes are clearly illustrated in
Figure~\ref{alpha_images}, where we show maps of the stellar and gas surface
densities.  In the gas surface density maps, the ISM shows significantly
different structure with EMF included.  With SN feedback alone, the ISM is
organised around a handful of large, feedback-driven voids in the inter-arm
regions of the galactic disc.  When EMF is included, the disc becomes more
uniform, with less prominent spiral arms and a more flocculent morphology.  The
stellar column density map illustrates qualitatively what we previously
quantified in Figure~\ref{alpha_clustering}.  EMF significantly reduces the
number of small, dense stellar groups, resulting in a disc that is smoother and
more uniform.  The handful of stellar groups that remain are found primarily
along spiral arms, with very few lying in either the interarm regions or the
outskirts of the stellar disc.

Differences in the spatial distribution of gas and stars, like those seen in the
face-on projections of Figure~\ref{alpha_images}, also exist in the vertical
structure of our simulated galaxies.  In Figure~\ref{alpha_scaleheight}, we show
the $z$-axis mass distribution of gas and stars in galaxies simulated with and
without empirically-motivated feedback.  As would be expected from the higher
outflow rates of the SN-only simulation, its vertical gas profile has much more
mass outside the gaseous disc ($|z|\ga1\kpc$) than the simulations with EMF.
Another interesting side-effect of the more vigorous, SN-stirred ISM is the
presence of stars at high galactic elevations in the SN-only simulation.  When
EMF is included, the stellar disc truncates completely at $|z|\sim2\kpc$.  When
this early feedback is omitted, we find broad wings of stellar mass, extending
even beyond $|z|\sim4\kpc$.  These stars may be kicked to higher orbits by
forming in gaseous regions with large $\sigma_z$, through dynamical
interactions, or may have formed in dense gas entrained along with outflows
\citep{Yu2020}.  Of course, it is important to realise that the stellar
population outside of the plane in real-Universe galaxies is mostly shaped by
satellite galaxy accretion \citep[e.g.][]{Helmi2018,Kruijssen2020,Naidu2021}.
Therefore, we leave further investigation of this phenomenon to a future study
that will include the effects of the cosmological environment on the shaping of
the halo star population.

In Figure~\ref{alpha_phase}, we show the impact of EMF on the physical state of
the gas in our simulated galaxies, visualised in the density-temperature plane.
The typical features of a three-phase ISM are seen both in the simulations with
and without EMF. A hot phase evolves adiabatically as it leaves the disc, a warm
ionized or neutral phase is found between $10^3{-}10^4$~K, which exists in rough
pressure equilibrium with a cool neutral phase between 10 and a few 100 K. As
the contours show, the cool phase is mostly isobaric ($\rho T \approx {\rm
const}$), while the warm phase shows a shallower slope, closer to (but not
fully) being isothermal. This shallow, pseudo-isothermal profile is simply a
result of the ionizing UV background.  The temperature and density histograms
along the horizontal and vertical edges of Figure~\ref{alpha_phase} clearly show
the quantitative differences between the simulations with only SN feedback and
those that include EMF.  As can be seen, the relative amount of hot, diffuse gas
is significantly higher for SN feedback alone than when EMF is included.  The
total mass of gas above $10^4$~K with SN only is $1.9\times10^9\Msun$, compared
to $\{3.9, 4.6\}\times10^8\Msun$ when including EMF with $\alpha=\{0.5, 1.0\}$.
This is another manifestation of the enhanced efficiency of SN feedback seen
previously in Figure~\ref{alpha_sfr}, driven by the enhanced clustering shown in
Figure~\ref{alpha_clustering}.  We verify that the changes we see in star
formation regulation and the ISM phase distribution are robust to galactic
stochasticity \citep{Keller2019} in Appendix~\ref{stochasticity}.

\subsection{Spatial de-correlation of gas and stars}

\begin{figure}
    \includegraphics[width=\hsize]{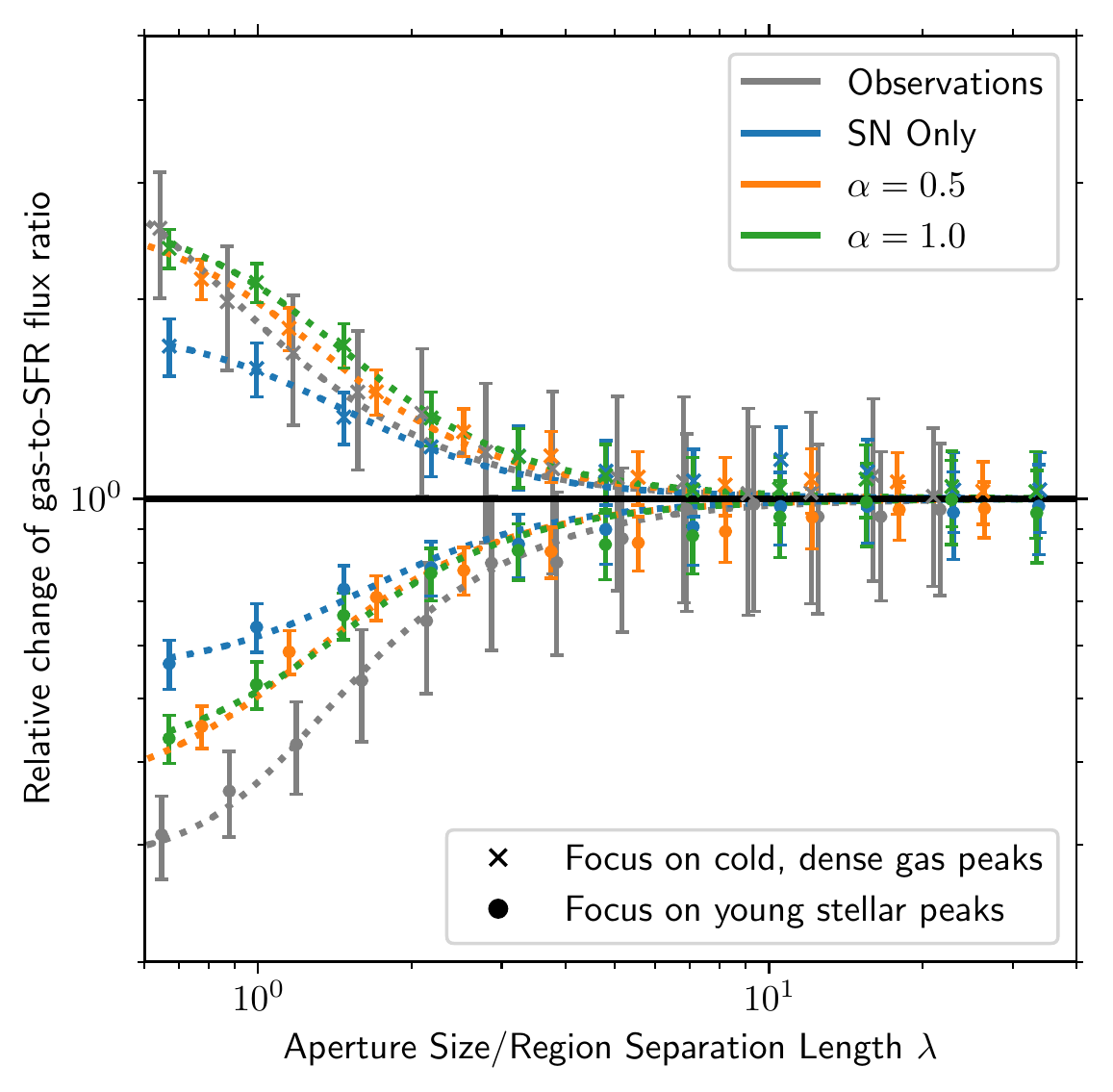}%
    \caption{Tuning fork diagram for observations of the LMC along with our
    three simulated galaxies.  The horizontal axis shows the aperture size over
    which gas and young stellar fluxes are measured, normalized to the inferred
    region separation length $\lambda$.  The vertical axis shows the ratio of
    the gas to young stellar (SFR) fluxes, normalized to the global average
    ratio.  The upper arm of the tuning fork shows measurements where apertures
    are centred on gas peaks, while the lower arm shows measurements where the
    apertures are centred on young stellar peaks.  As can be seen, the
    simulations with SN feedback alone show a smaller, more flattened opening on
    small scales, while simulations that include EMF show an opening comparable
    to the tuning fork observed for the LMC \citep{Ward2020,Kim2021}.}
    \label{alpha_tuning}
\end{figure}

\begin{table}
\centering
    \begin{tabular}{|l|r|r|}
        Galaxy & $t_{\rm gas}$ (Myr) & $t_{\rm FB}$ (Myr)  \\
        \hline
        \hline
        LMC (observations) & $11.1^{+1.6}_{-1.7} $ & $1.1^{+0.3}_{-0.2}$ \\
        PHANGS++ (observations) & $19.8\pm6.1$ & $3.3\pm1.2$ \\
        SN Only & $14.8^{+1.7}_{-1.5}$ & $3.3^{+0.7}_{-0.7}$ \\
        EMF $\alpha=0.5$ & $10.7^{+0.9}_{-0.8}$ & $2.0^{+0.4}_{-0.6}$ \\
        EMF $\alpha=1.0$ & $9.3^{+1.3}_{-0.8}$ & $1.4^{+0.6}_{-0.5}$ \\
        \hline
    \end{tabular}
    \caption{Time-scales measured using {\sc Heisenberg} for observations of the
    LMC, the averaged PHANGS+NGC300+NGC5194 (PHANGS++ hereafter) observations used as inputs, and for our three
    simulated galaxies, using gas column density maps for gas with $n>100\hcc$.
    Uncertainties for the PHANGS++ observations are the standard deviations
    of the sample, while uncertainties for the individual galaxies (LMC and
    simulations) are calculated using the {\sc Heisenberg} analysis code.
    As can be seen, both the cloud lifetimes $t_{\rm gas}$ and feedback
    timescales $t_{\rm FB}$ for galaxies simulated with EMF are within the
    observational uncertainties of clouds within the LMC, but shorter than the
    timescales from the PHANGS++ galaxies.}
    \label{alpha_heisenberg}
\end{table}

\begin{figure}
    \includegraphics[width=\hsize]{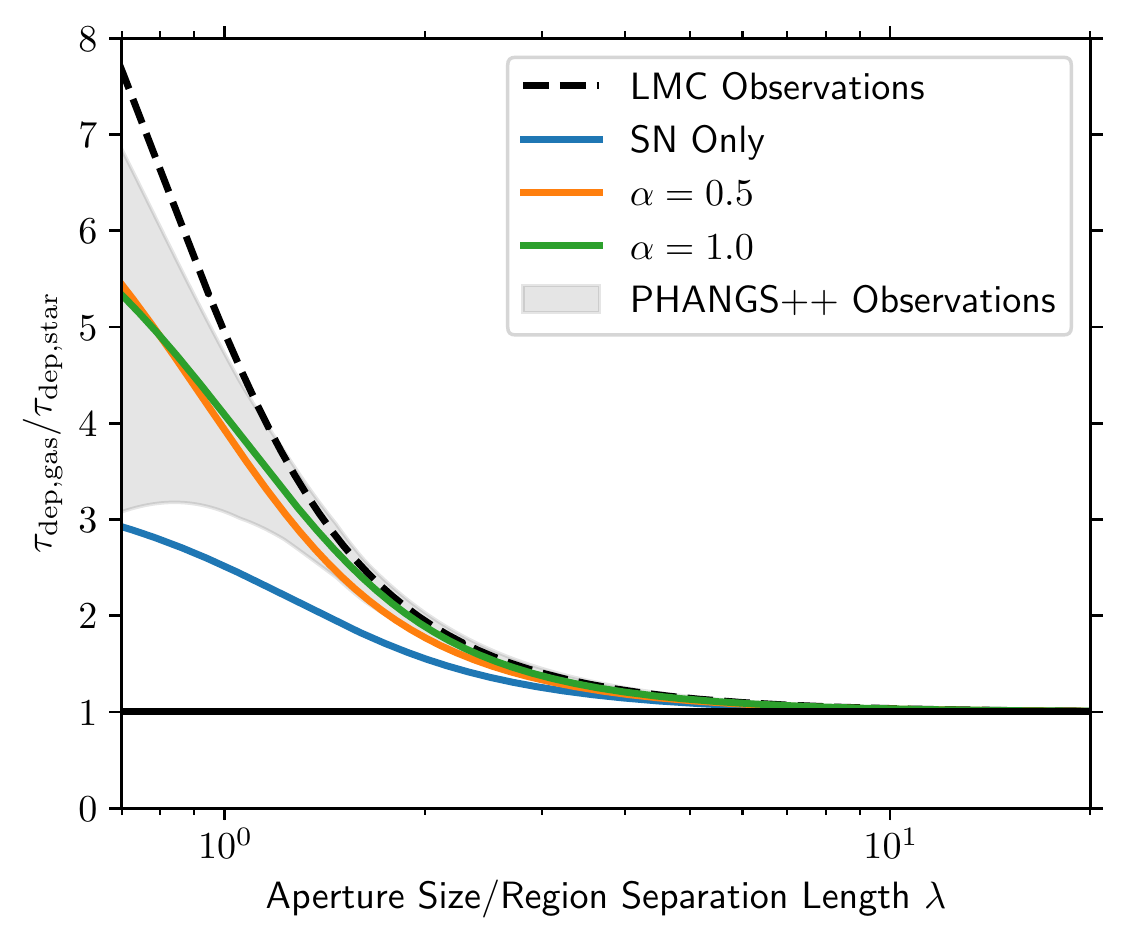}%
    \caption{De-correlation ratio of gas and stars at different aperture sizes,
    for observed galaxies (shaded region and black dashed curve) and our simulations (coloured lines).
    The de-correlation is measured by the ratio of depletion times measured in
    apertures centred on dense gas peaks and those measured in apertures centred
    on young stellar peaks, i.e.\ the ratio between both branches in
    Figure~\ref{alpha_tuning}.  As can be seen, the de-correlation measured by
    $\tau_{\rm dep,gas}/\tau_{\rm dep,star}$ for galaxies simulated with EMF
    lies within the $\pm 1\sigma$ range of observed galaxies, shown by the grey
    shaded region.  With SN feedback only, this ratio is reduced by a factor of
    $\sim2$, and no longer agrees with the observed range of values.}
    \label{alpha_decorr}
\end{figure}

The observationally-derived quantities we have used for our EMF simulations were
determined by \citet{Kruijssen2019b} and \citet{Chevance2020,Chevance2022}, who
analysed the spatial de-correlation of gas and stars down to $\la100$~pc scales
in 10 nearby galaxies, using the {\sc Heisenberg} code \citep{Kruijssen2018}.
In this section, we analyse our simulated galaxies with the same methodology, to
reveal how EMF changes the spatial de-correlation of star-forming gas and young
stars.  In order to do this, we generate mock surface density maps of
star-forming gas and young stars.  For the gas maps, we use all gas cells with
densities above our star formation threshold, $100\hcc$.  For the young stars,
we select all stars with ages below $10\Myr$.  We then generate $20\times20\kpc$
column density maps, with a resolution of $1000\times1000$ pixels (such that
each pixel corresponds to $20\pc$). We then smooth the images with a Gaussian
beam with $\sigma=20\pc$ (or a full width half maximum of $\sim50\pc$),
comparable to the size of the resolution element of the observations analysed by
\citet{Kruijssen2019b} and \citet{Chevance2020,Chevance2022}.  These maps are
then stored as {\sc FITS} files to be read as input by the {\sc Heisenberg}
code.  

In Figure~\ref{alpha_tuning}, we show the key qualitative metric produced by
{\sc Heisenberg}, the so-called ``tuning fork'' diagram that is the fundamental
relation defined by the \citetalias{Kruijssen2014a} ``uncertainty principle for
star formation''.  This tuning fork shows the change in gas depletion times
$\tau_{\rm dep}=\Sigma_{\rm gas}/\Sigma_{\rm SFR}$ in apertures of various size
relative to the globally-averaged depletion time.  These apertures are centred
on flux peaks identified in the column density maps that we have generated.  The
upper branch, where we focus apertures on gas peaks, rises above the global
average depletion time.  The lower arm, where we focus on stellar peaks, dips
below the global average depletion time.  The width of this opening is
determined by the overlap time ($t_{\rm FB}$), i.e.\ the duration for which we
expect to see spatially-correlated gas and stellar flux.  As $t_{\rm FB}$ goes
to zero, the opening widens because cold, dense gas is instantaneously removed
from the environment of young stars -- observing young stars then implies
\textit{not} observing gas, and therefore measuring a short depletion time.  As
$t_{\rm FB}$ becomes larger, the spread of this opening shrinks and the branches
of the tuning fork flatten, because young stars and star-forming gas co-exist
for a longer duration of a cloud's star forming lifetime.  The tuning fork will
not flatten completely as long as the cloud lifetime is longer than the overlap
time-scale $(t_{\rm gas}>t_{\rm FB})$ and there exist some clouds without young
stars (and vice versa).

In contrast to the feedback time-scale $t_{\rm FB}$, the ratio of cloud lifetime
$t_{\rm gas}$ and young stellar lifetimes control the vertical asymmetry in the
tuning fork, and the average region separation length $\lambda$ controls where
the tuning fork begins to open (for more on the information that can be gleaned
from the tuning fork diagram, see the detailed description in sect. 3.2.11 of
\citealt{Kruijssen2018}).  We normalize our tuning fork diagram by the region
separation length $\lambda$ in order to focus exclusively on the relative width
of the opening, which shows the spatial de-correlation of gas and stars and
probes the quantities of interest, i.e.\ the duration of the feedback time-scale
$t_{\rm FB}$ and the specific terminal momentum $p_0$.  In addition to the three
simulated galaxies, we also show a tuning fork derived from observations of the
Large Magellanic Cloud (LMC; \citealt{Ward2020,Kim2021}). We have explicitly not
included these measurements in our calculation of the input parameters for EMF,
allowing us to use these observations as an independent test of the spatial
de-correlation in our simulated galaxies. We select the LMC, because its tuning
fork and underlying time-scales provide the best match to those obtained for the
simulations analysed here.  As can be seen, the de-correlation between gas and
stars in the simulation without EMF is too low, resulting in a tuning fork
flatter than both the EMF simulations and the observed LMC results.

In Table~\ref{alpha_heisenberg}, we show the two primary time-scales derived
using {\sc Heisenberg}, the gas cloud lifetime $t_{\rm gas}$ and the
overlap/feedback time-scale $t_{\rm FB}$ for our simulated galaxies, as well as
for the observations of the LMC \citep{Ward2020,Kim2021} and the PHANGS++ sample.  As these data show,
the LMC and our EMF simulated galaxies have comparable cloud lifetimes,
approximately within $1\sigma$ of each other, with the EMF simulated galaxies
having slightly shorter (but mutually indistinguishable) lifetimes of $t_{\rm
gas}\approx10\Myr$, compared to $t_{\rm gas}\approx15\Myr$ for galaxies
simulated with SN alone.  The feedback time-scales are also significantly higher
(as we would expect) in the SN-only simulated galaxy, more than $3\sigma$ above
the feedback time-scale measured in the LMC. By contrast, the feedback
time-scales inferred for the simulations with EMF are relatively insensitive to
$\alpha$, with values ranging from $0.5\sigma$ higher for $\alpha=1.0$ to
$1.3\sigma$ higher for $\alpha=0.5$.

While the results we find are consistent with a galaxy that was not included
among the data we use to derive the parameters $p_0$ and $t_{\rm FB}$, we find
that the simulated galaxies here do not match the timescales for the observed
PHANGS++ galaxies used to derive these parameters.  The cloud lifetimes $t_{\rm
gas}$ and feedback timescales $t_{\rm FB}$ are both found to be lower in our
simulated galaxies that include EMF compared to the median observations of the
PHANGS++ sample (though still well within the observed range).  In order to determine
the sensitivity of these derived values to parameters other than the feedback
prescription, we generated gas column density maps with a lower gas cell density
threshold ($30\hcc$), and re-calculated $t_{\rm gas}$ and $t_{\rm FB}$. In
Appendix~\ref{sfparams}, we also examine the sensitivity to parameter choices
for star formation model used.  We find  we derive
significantly $(\sim 4\times)$ longer cloud lifetimes, as well as somewhat
longer feedback timescales for all cases when lower-density gas maps are used
(as has been seen in the observations of the atomic gas cloud lifetimes seen in
\citealt{Ward2020}).

It should also be noted that while our galaxies simulated with EMF match well
the observed feedback timescales in the LMC, there is significant scatter in the
PHANGS++ sample, with feedback timescales ranging from $1.0-4.8\Myr$.
Determining the origin of this scatter, and the potential dependence on
environment (see for example \citealt{Chevance2022}), will require further
observational studies.  However, in order to fully self-consistently
compare measurements of $t_{\rm gas}$ and $t_{\rm FB}$ from simulated galaxies,
a full treatment of radiative transfer (RT) and CO chemistry to derive mock CO
and $H\alpha$ maps to directly match the observational quantities used to derive
these timescales \citep{Fujimoto2019}.  While this is beyond the scope of this
paper, we are now working on a careful examination of how to most consistently
compare RT mock observations of simulated galaxies to true observations (Petkova
et al. in prep).

An insightful, alternative quantification of the spatial de-correlation between
gas and young stars can be seen in Figure~\ref{alpha_decorr}.  There we show the
ratio of the upper branch (centred on dense gas peaks) to the lower branch
(centred on young stellar peaks), fitted using {\sc Heisenberg}, which we refer
to as the de-correlation ratio.  This corresponds to the ratio of depletion
times measured centred on dense gas versus young stellar peaks, and quantifies
the width of the tuning fork in logarithmic space. Importantly, this width is
controlled entirely by the feedback time-scale $t_{\rm FB}$, and the quantity
shown in Figure~\ref{alpha_decorr} therefore allows us to isolate the
contribution of $t_{\rm FB}$ only, without needing to worry about the other
dependences on the cloud lifetime and the region separation length.  For
reference, we also show the $\pm 1\sigma$ range of the de-correlation ratio
observed in the PHANGS++ sample of nearby galaxies used to determine the EMF parameters
\citep{Kruijssen2019b,Chevance2020,Chevance2022}.  As can be seen, the
de-correlation ratio for each of our galaxies simulated with EMF lies within the
$\pm 1\sigma$ scatter of the observed PHANGS++ galaxies, while the galaxy
simulated with SN feedback only lies significantly below this range.  This
directly reflects the longer feedback time-scale of the SN-only simulation, and
clearly demonstrates that EMF is able to reproduce the spatial de-correlation
observed in local spiral galaxies, while feedback from SN alone cannot.

\begin{figure}
    \includegraphics[width=\hsize]{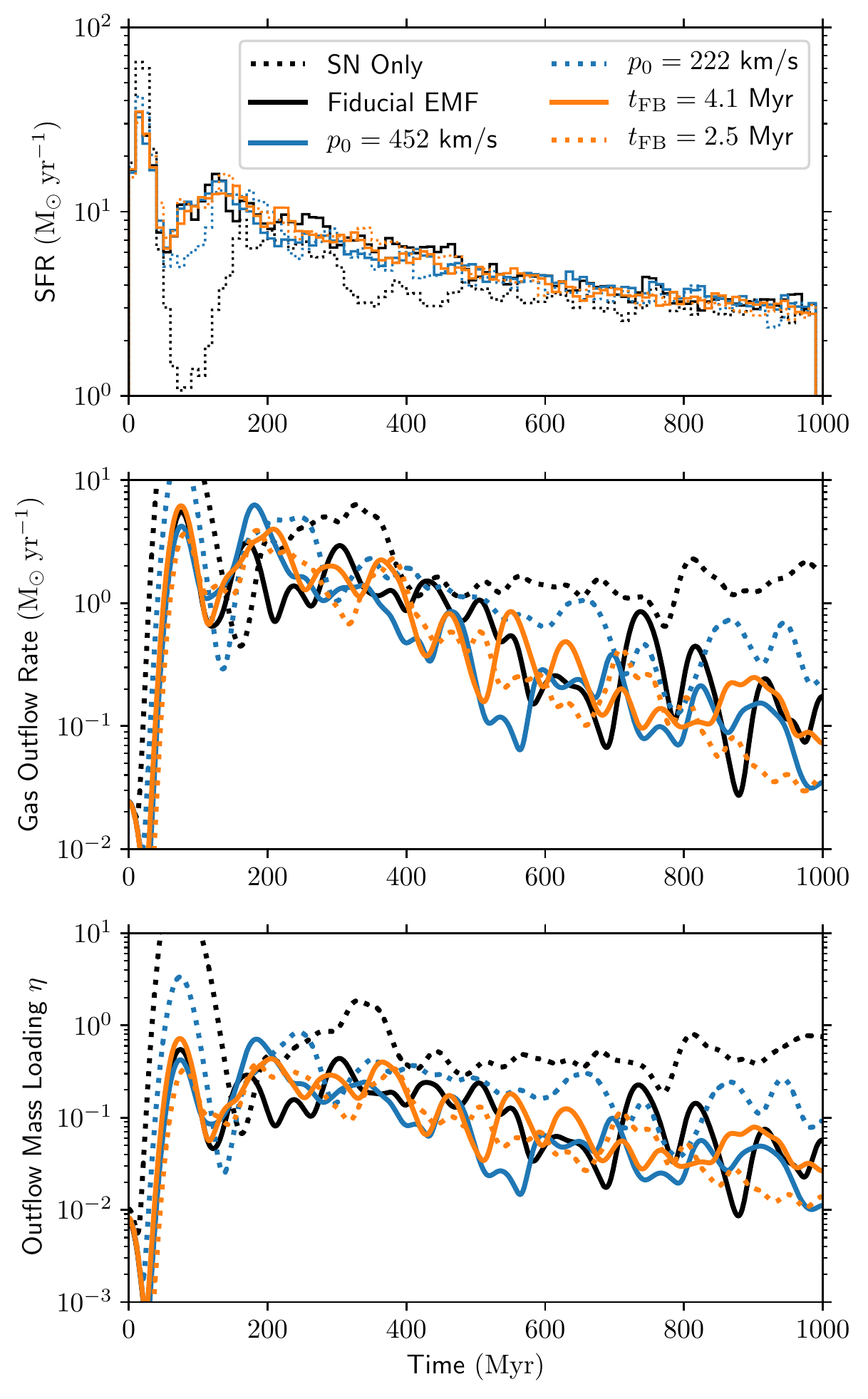}%
    \caption{Star formation rates (top panel), outflow rates (middle panel), and
    mass loadings (bottom panel) for EMF with 25$^{\rm th}$-to-75$^{\rm th}$
    percentile variations in the observationally-derived parameters $p_0$ (blue
    curves) and $t_{\rm FB}$ (orange curves).  Fiducial parameters are shown in
    black for the SN-only run (solid) and EMF run with $\alpha=1.0$ (dotted).
    As can be seen, varying $t_{\rm FB}$ between $2.5\Myr$ and $4.1\Myr$ has
    little impact on the star formation or outflow rates.  However, if $p_0$ is
    reduced to $222\kms$ (from the median observed value of $375\kms$), we see
    slightly higher mass loadings and outflow rates, as expected.  An analogous
    shift is not found when increasing $p_0$ to $452\kms$, because this does not
    significantly reduce the overall outflow rates.}
    \label{param_sfr}
\end{figure}
\begin{figure}
    \includegraphics[width=\hsize]{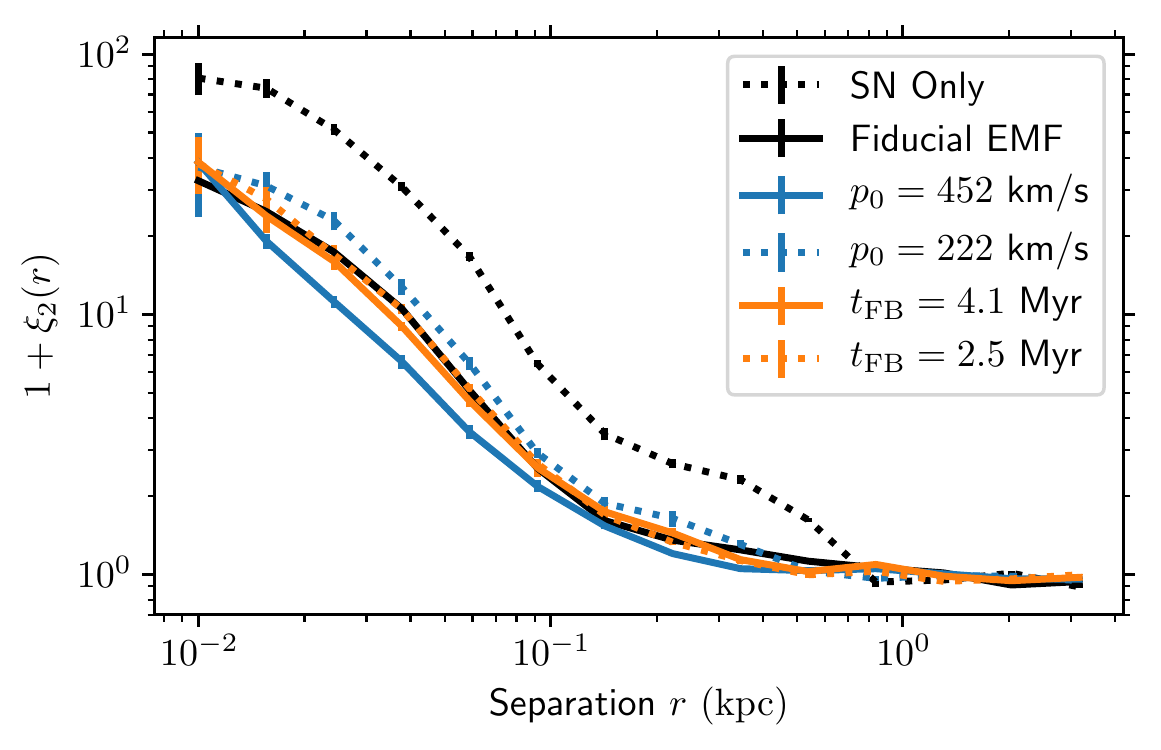}%
    \caption{Two-point correlation function of young ($<20\Myr$) stars for
    25$^{\rm th}$-to-75$^{\rm th}$ percentile variations in $p_0$ and $t_{\rm
    FB}$. As can be seen, neither increasing nor decreasing $t_{\rm FB}$
    produces a change in stellar clustering larger than the uncertainties on
    $\xi_2$.  Increasing $p_0$ produces a slight decrease in clustering on
    intermediate ($30-100\pc$) scales, but no detectable difference on the
    smallest scales ($<20\pc$).  Decreasing $p_0$ pushes the results closer to
    those seen with SN feedback alone, raising the smallest-scale correlation
    function $\xi_2$ by a factor of $\sim2$.}
    \label{param_clustering}
\end{figure}

\subsection{Varying the observationally-derived parameters}

In the previous sections, we have examined the effect of varying the
observationally unconstrained expansion exponent $\alpha$, and found that for
physically reasonable values ($\alpha=0.5-1.0$), galaxies simulated with EMF
show little sensitivity to $\alpha$.  While the two other parameters $p_0$ and
$t_{\rm FB}$ are derived using observational data, there is non-trivial scatter
in these observed quantities.  Irrespectively of whether this is due to
environmental variations in the fundamental feedback processes that we
parameterise, or simply because of observational uncertainty, it is useful to
know how variations in $p_0$ and $t_{\rm FB}$ change the overall impact of EMF
on star formation regulation in our simulated galaxies.  In this section, we
show how star formation regulation is affected when varying $p_0$ from $222\kms$
to $452\kms$ and $t_{\rm FB}$ from $2.5\Myr$ to $4.1\Myr$, which corresponds to
the interquartile ranges of the observational measurements of these quantities
(see Section~\ref{obs_measurements}).  We perform four additional simulations in
which we fix $\alpha=1.0$, while varying $p_0$ and $t_{\rm FB}$ independently of
each other. For runs with modified $p_0$, we keep $t_{\rm FB}$ fixed at
$3.3\Myr$, while for modified $t_{\rm FB}$, we keep $p_0$ fixed at $375\kms$.

In Figure~\ref{param_sfr}, we show how the star formation and outflow rates are
changed by modifying $p_0$ and $t_{\rm FB}$.  As we might expect, lowering $p_0$
pulls the results towards what we see from SN feedback alone (as we are reducing
the overall effectiveness of the early stellar feedback).  While the SFRs show
little difference for any of the cases we examine, the outflow rates and mass
loadings do show a small but significant difference.  Comparing the outflow
rates for the variations in $p_0$, the mean outflow rates over the final
$500\Myr$ for $p_0=\{222,375,452\}\kms$ are $\{0.58,0.20,0.12\}\Msun\yr^{-1}$.
Changes in $t_{\rm FB}$ do not produce as significant a change compared to $p_0$
(as we might expect).  Adopting $t_{\rm FB}=\{2.5,3.3,4.1\}\Myr$ produces mean
outflow rates of $\{0.12,0.20,0.16\}\Msun\yr^{-1}$.  

As in Figure~\ref{alpha_clustering}, we can see in Figure~\ref{param_clustering}
that EMF reduces the small-scale ($<1\kpc$) clustering of young stars in our
simulated disc galaxy.  Increasing the value of $p_0$ slightly decreases the
stellar clustering on scales of $20-100\pc$, but a much larger increase in the
two-point correlation function on small scales is seen when reducing $p_0$ to
$222\kms$.  In turn, this increase in small-scale clustering drives the larger
outflow rates and mass loadings we have seen in Figure~\ref{param_sfr}.  This
weaker form of early feedback produces a result intermediate between EMF with
our fiducial parameters and the SN-only example.  By contrast, all variations of
$t_{\rm FB}$ are within the error bars of each other, and are also consistent
with the fiducial parameter choices.  We likely would need to vary $t_{\rm FB}$
well beyond the observational constraints to see a significant change in the
clustering of young stars or the outflow mass loadings.

\section{Discussion}
\label{discussion}

\subsection{Previous models for early feedback in galaxy simulations}
Early stellar feedback has been studied extensively in simulations of both
isolated galaxies and cosmological zoom-in simulations
\citep[e.g.][]{Wise2008,Stinson2013,Rosdahl2015}.  Many different models have
been proposed, and there is still considerable uncertainty as to the
quantitative impact of different physical processes and assumptions, as well as
different numerical approximations of the same feedback processes. 

A number of studies have taken an agnostic approach to the specific driving
mechanisms of early stellar feedback, as we have done here.  Rather than
explicitly modelling individual early stellar feedback processes, these studies
use simplifying assumptions about the total stellar feedback budget as the
foundation of a model for early stellar feedback.  The ``MaGICC'' early feedback
model first presented in \citet{Stinson2013} simply injects $10\%$ of the total
stellar luminosity as thermal energy, and has been applied to the large
cosmological zoom-in simulations of over 100 galaxies in NIHAO \citep{Wang2015}.
\citet{Stinson2013} showed in cosmological zoom-in simulations of an $L^*$
galaxy that this simple model for early stellar feedback could significantly
reduce the overall stellar mass of the galaxy, and help to produce a
disc-dominated system with a flat rotation curve.  In the recent study by
\citet{Semenov2021}, stellar winds were approximated by simply shifting the
time-scale over which SN detonate to begin $0\Myr$ after a star particle forms.

A tremendous amount of effort has been spent developing models for early stellar
feedback that rely on ionizing radiation.  As we detailed in
Section~\ref{FB_mechanisms}, ionizing radiation can disrupt star-forming gas
through direct and indirect radiation pressure, photoionization, and by the
expansion of overpressured HII regions \citep{Kim2018c}.  Studies looking at
early radiative feedback have typically used one of two approaches: capturing
radiative feedback through explicit radiative transfer, or by approximating each
mechanism of radiative feedback with sub-resolution models.  Explicit radiative
transfer tends to be numerically expensive, and so approximations are required
to achieve reasonable performance in both the scaling of the algorithm with
source and particle number as well as the timestep required for numerical
stability.  Early studies employing ray-tracing techniques
\citep[e.g.][]{Wise2008} were limited to studying extremely high-redshift
objects, while others looked at smaller, dwarf-scale objects
\citep[e.g.][]{Kim2013}.  The development of tree-based radiative transfer
algorithms \citep{Kannan2014,Obreja2019a,Benincasa2020} may allow much
larger-scale radiative transfer simulations in the future.  Even when the full
radiative transfer equations (including scattering and re-emission) are
followed, limited resolution in the highest density regions of the ISM may still
require significant sub-grid approximations \citep{Rosdahl2015}.  This has been a
particular focus for studies of reionization, as the escape fraction of ionizing
photons depends strongly on the fractal, porous structure of unresolved GMCs
\citep{Kimm2014,Ma2015,Trebitsch2017}.

Sidestepping the additional cost of full radiative transfer has generally relied
on the assumption that most of the impact of radiative feedback is local,
concentrated in the immediate vicinity of a star-forming region.  Local
radiative models have been built to capture the effects of HII region expansion
\citep{Jeffreson2021b}, radiation pressure \citep{Roskar2014,Ceverino2014}, and
photoionization \citep{Smith2021}. In general, these models find that radiative
feedback is sub-dominant to SN feedback in most galactic environments
\citep{Su2017}, but may act to either boost \citep{Hopkins2011} or suppress
\citep{Smith2021} SN-driven galactic outflows.  Further research is required to
fully understand the interplay of radiative early feedback, SN driven outflows,
and galaxy-scale regulation.

An exciting new field of study for the comprehensive effects of early stellar
feedback is in ultra-high resolution simulations of isolated dwarf galaxies,
where the internal structure of star-forming GMCs becomes resolvable
\citep{Emerick2019,Hu2017,Smith2021}.  Both \citet{Hu2017} and \citet{Smith2021}
find similar effects in these well-resolved simulations when photoelectric
heating and photoionization are included on top of SN feedback alone.  They find
increased outflow rates and a significantly more disrupted ISM structures when
early feedback is omitted. This agrees qualitatively with the results we show
here, even though their simulations have mass resolutions $\sim 1000$ times
better than the galaxies we have simulated (note that their simulated galaxies
are also roughly a factor of $\sim 100$ less massive than the $L^*$ objects we
simulate here). With baryonic mass resolutions of $\ll 100\Msun$, not only is
the internal structure of GMCs accessible, but simulations become capable of
tracking individual massive stars, forming physically-resolved (although not
dynamically-resolved) star clusters and resolving the complex interplay between
feedback fronts driven by each massive star.  These simulations offer a new
frontier to understand how feedback disrupts these clouds in a realistic
galactic environment.  Unfortunately, these extreme resolutions are well out of
reach for simulations of more massive galaxies, especially those that include a
full cosmological environment.  Despite this, they are an excellent tool for
understanding how the observational momentum inputs and time-scales arise, and
what feedback mechanisms dominate this momentum budget.

In addition to detailed studies looking at individual early feedback processes,
there is an abundance of studies that attempt to include complete accountings of
the stellar feedback budget, including both winds and radiative processes
\citep[e.g.][]{Trujillo-Gomez2015,Goldbaum2016,Marinacci2019}.
\citet{Agertz2013} presented a detailed budgeting of the momentum and energy
injection from massive stars, and developed a model that has been used to study
the cosmological evolution of galaxies \citep{Agertz2015}.  The FIRE
\citep{Hopkins2014} and FIRE-2 \citep{Hopkins2018b} have been used for numerous
studies using cosmological zoom-in simulations of galaxies with virial masses
ranging from $\sim10^9\Msun$ to $\sim10^{12}\Msun$.  FIRE includes explicit
sub-grid models for early feedback from stellar winds, radiation pressure, and
photoionization.  These early feedback mechanisms were shown in
\citet{Hopkins2014} to enhance the efficiency of SN at regulating the stellar
and baryonic mass of galaxies, though there remains some uncertainty as to the
magnitude of the effect in light of the approximate nature of the sub-grid
models used. 

Physically-motivated models for individual feedback mechanisms will be a
fruitful comparison tool to the EMF model we have presented here.  While EMF is
designed to match observed cloud-scale feedback timescales, we have taken an
agnostic approach as to which particular feedback mechanism is dominant, whether
any single mechanism can explain these observed timescales, and if this depends
on galactic environment.  By comparing different physical mechanisms of early
stellar feedback to both observations and simulations using EMF, we can better
understand the underlying physics that drives the disruption of GMCs in the
diverse galactic environments they find themselves in.

\subsection{Spatial de-correlation and tuning forks in previous simulations}
Since the introduction of the ``tuning fork diagram'' by \citet{Schruba2010},
its analytical characterisation in terms of time-scales by
\citetalias{Kruijssen2014a}, and the initial suite of galaxy simulations
demonstrating its suitability for accurately characterising the cloud lifecycle
\citep{Kruijssen2018}, five further simulation studies have used this technique
to examine the lifetimes of molecular clouds and the cloud-scale feedback cycle.
While these studies did not attempt to build feedback models directly from
observational measurements of cloud life cycles, they did use the tuning fork
diagram as a diagnostic for characterising stellar feedback processes.

The first of these studies, \citet{Fujimoto2019}, used AMR simulations of an
isolated, Milky Way-like galaxy similar to what we have examined here, but with
higher resolution refinement to a minimum cell size of $\sim7\pc$ in the
Eulerian hydrodynamic code {\sc ENZO} \citep{Bryan2014}.  Their star formation
algorithm is the same as what we use here, and they include stellar feedback
from SNe and photoionization from young massive stars.  Their photoionization
early feedback model calculates the Str\"omgren volume from the luminosity of
stars, and then heats an appropriate region to $10^4\K$.  For SNe feedback,
their simulations directly inject the terminal momentum ($5\times10^5\Msun\kms)$
into the cells surrounding each SN event, and deposit the remaining thermal
energy in the SN host cell. The \citet{Kimm2014} mechanical feedback model that
we apply deposits comparable momentum in the limit of infinitely poor
resolution, but additionally adaptively scales the ratio of kinetic to thermal
energy injected based on the local resolution.  Unlike our study,
\citet{Fujimoto2019} use the {\sc Despotic} \citep{Krumholz2014} radiative
transfer code to post-process their simulations, in order to generate CO
$J=1\rightarrow0$, HI, and H$\alpha$ emission maps.  They find that the feedback
model in their simulations produced long-lived clouds, with average lifetimes of
$36_{-6}^{+4}\Myr$ and extremely long feedback time-scales of $23\pm1\Myr$,
measured using the \citetalias{Kruijssen2014a} methodology.
\citet{Fujimoto2019} attributes these long lifetimes to ineffective
photoionization feedback.  These simulations seem to regulate star formation
through slow and (very) inefficient star formation, rather than the fast and
inefficient mode seen in our simulations and in the observations of
\citet{Kruijssen2019b}, \citet{Chevance2020,Chevance2022}, and \citet{Kim2021}.

Another recent trilogy of papers
\citep{Jeffreson2020,Jeffreson2021a,Jeffreson2021b} have examined tuning fork
diagrams in simulated Milky Way-like galaxies.  These simulations use an
identical model for SNe feedback as we have employed here (in fact, the same
implementation in the {\sc Arepo} numerical code), but also include a model for
momentum injection via unresolved HII regions described in
\citet{Jeffreson2021b}.  \citet{Jeffreson2021b} finds that the small-scale
spatial de-correlation of gas and stars is impacted somewhat less significantly
by early feedback from HII regions, reducing $t_{\rm FB}$ from $>5\Myr$ to
$4.4\Myr$. In \citet{Jeffreson2021a}, the authors use high-resolution
simulations of a Milky-Way like galaxy, along with the {\sc Astrodendro}
\citep{Robitaille2019} cloud-finding tool to trace individual molecular clouds
and quantify their evolution over time.  They find that in these galaxies, the
opening of the tuning fork on small scales gives de-correlation ratios
$\tau_{\rm dep,gas}/\tau_{\rm dep,star}\sim4$, within the range of the PHANGS
observations, although somewhat lower than the values of $\sim5.5$ we find here.
The authors then use these same simulations to a derive new scaling relations
for cloud lifetimes in Milky Way-like galaxies.  They find the average
time-scales derived from directly tracking clouds from high-cadence simulations
($1\Myr$ temporal resolution) is generally in agreement with the averaged cloud
lifetimes derived using the methodology of \citet{Kruijssen2018}.

More recently, a comprehensive study by \citet{Semenov2021} has looked at the
combined effects of explicitly modelled radiative transfer, H$_2$ chemistry,
unresolved turbulent pressure, early mechanical feedback, and a turbulence-based
star formation model.  They model early feedback through UV photoheating and
photoionization, along with a simple model for early mechanical feedback that
simply shifts the injection of SN energy to begin immediately after the first
star forms (a discussion of the impact of SN injection timings can be found in
\citealt{Keller2022a}).  Unresolved turbulence is captured using a model first
presented by \citet{Semenov2016}, which accounts for the turbulent energy
cascade below the resolution scale of the simulations.  \citet{Semenov2021}
examine a number of star formation models, including a simple Schmidt law
similar to what we have used here, a Schmidt law with a virial parameter cutoff
mimicking the FIRE-2 star formation model \citep{Hopkins2018b}, and a model
based on the \citet{Padoan2012} small-scale simulations of star formation (a
similar model was also recently described in \citealt{Gensior2020}).  They use
initial conditions of an isolated disc galaxy with general properties designed
to match NGC300, allowing them to compare to the observations presented by
\citet{Kruijssen2019b}.  With an array of simulations using different selections
of feedback processes and star formation parameter choices, they demonstrate how
strong early feedback increases the de-correlation ratio, as we have
demonstrated here for the EMF model.  For very high star formation efficiencies
with a Schmidt-type star formation law, they in fact find de-correlation ratios
much higher than NGC300.  In their simulations with SN feedback alone, they also
find that the de-correlation ratio is much smaller than the observational values
(as indicated by a smaller opening of the tuning fork on small scales),
consistently with our results.  Interestingly, they also show that
self-consistent radiative early feedback is important for setting the region
separation length $\lambda$.

\subsection{Limitations and future directions}
The EMF model we present here relies primarily on the observational constraints
on $p_0$ and $t_{\rm FB}$.  Currently, we have used a sample only 10 nearby disc
galaxies to obtain these quantities.  Future observations of a larger galaxy
sample (J.~Kim et al.\ in prep.) will provide us with the statistics needed to
improve our understanding of how these parameters change in different
environments. This will be aided further by additionally extending the sample to
more extreme environments, such as metal-poor dwarf galaxies, gas-rich
starbursts, and massive ellipticals.  Of particular interest will be to derive
scaling relations between $p_0$, $t_{\rm FB}$, and the galactic environment
(e.g.\ local ISM properties).  Further, more detailed studies of the
relationship between the ISM and feedback timescales in observed galaxies will
allow us to develop an improved, environmentally-dependent version of EMF that
will more accurately model the effects of early feedback in galaxies very
different than the isolated $L^*$ discs at $z=0$ that we have studied here.  

Of particular interest and challenge is the pressure (and thus)
scale height of the ISM.  As we derive our model using self-similar, spherical
wind models, a feedback bubble will diverge from the simple
equation~\ref{selfsim} when the ISM pressure becomes comparable to the ram
pressure of the feedback front, and when the radius of the bubble becomes
comparable to the ISM scale height \citep{MacLow1988}.  Treating the
non-spherical evolution of non-self-similar feedback fronts is a particularly
challenging problem for sub-grid models for feedback, especially if one wishes
to build models with reasonable convergence properties.  In particular, bridging
the resolution gap from low-resolution simulations that do not resolve
the ISM scale height (as is the current state of the art in large cosmological
volumes) to higher resolution simulations of galaxies with a well-resolved ISM
may not be possible with a single model for feedback.

Environmentally-dependent EMF will be of particular importance for cosmological
simulations, as even galaxies that are similar to the isolated discs we study
here at $z=0$ will spend much of their lifetime as considerably smaller, more
metal-poor progenitors in a more merger-rich environment.  EMF is well-suited
for cosmological simulations, as it is numerically inexpensive, and shows good
convergence properties (see Appendix~\ref{convergence}).  Future observations
will help reveal how generalizable the current values for $p_0$ and $t_{\rm FB}$
are for both the progenitors of $L^*$ galaxies as well as for a wide variety of
galaxy masses observed at $z=0$.

We have demonstrated that EMF can reproduce the observed spatial de-correlation
of gas and stars, and presented a method for deriving the terminal momentum
injected by early feedback from this observed de-correlation.  However, this
model is incapable of investigating the ultimate cause of the observed short
feedback time-scales, as well as how the small-scale interaction of multiple
feedback processes produce the momentum inferred from observational data.  The
values of $p_0$ and $t_{\rm FB}$ are assumed as input parameters to our model,
and are agnostic to the underlying physics that set these quantities.  High
resolution simulations that resolve individual massive stars and the GMCs in
which they are born are the most effective way forward to answer this question,
whether that be simulations of individual GMCs \citep[e.g.][]{Geen2018},
stratified slices of the ISM \citep[e.g.][]{Gatto2017}, or isolated dwarf
galaxies \citep[e.g.][]{Smith2021}.

\section{Conclusions}
\label{conclusions}
We have presented here a new numerical model for early stellar feedback that for
the first time uses parameters directly constrained through observational
measurements.  Using the self-similar evolution of feedback fronts, we apply
measurements of the feedback timescale $t_{\rm FB}$, cloud radius $r_{\rm cl}$,
and star formation efficiency $\epsilon_{\rm SF}$ to derive the
empirically-motivated terminal momentum $p_0$.  Our model captures uncertainty
in the cloud-scale star formation rate and the true driving mechanism through a
single dimensionless quantity, the expansion exponent $\alpha$.  As we have
shown, not only is this parameter physically constrained to a narrow range,
$\sim0.5-1$, and galaxies simulated using empirically-motivated early stellar
feedback are also relatively insensitive to the value of the expansion exponent.

Using a suite of simulated Milky Way-like galaxies, we have examined how
observationally-constrained early stellar feedback drives both quantitative and
qualitative changes in the overall evolution of these simulated galaxies. We
find that the overall regulation of the star formation rate is only marginally
affected by the inclusion of EMF.  After an initial period of settling, galaxies
simulated with EMF show a slightly higher ($\sim10-15\%$) star formation rate,
and a slightly steeper Kennicutt-Schmidt relation, than those simulated with SN
feedback alone.  While an increase in star formation due to additional feedback
may seem surprising at first, we find that this occurs due to a fundamental
change in the spatio-temporal distribution of star formation.  Without the early
disruption of star forming clouds provided by EMF, we see a significant,
quantitative increase in star particles forming co-spatially.  This in turn
results in a marked change in how SNe couple to the surrounding ISM.  Without
EMF, overlapping SNe blastwaves generate more hot, diffuse gas within the ISM,
losing less of their energy to radiative losses.  Qualitatively, this produces
an ISM with large voids, as seen in column density maps when EMF is omitted.
Quantitatively, the reduction of hot, SN-heated gas we see with EMF results in a
$\sim1$ dex decrease in outflow mass loadings, and produces thinner gaseous and
discs.  This suggests that EMF may be a solution to the difficulty recent
simulation studies \citep{Roskar2014,Grand2017} have found in producing truly
thin discs, though this will require fully cosmological simulations to fully
explore.  The significant changes in the SN-driven outflow behaviour that EMF
produces also has significant implications for future observational surveys of
outflow launching and the resolved mass and metal content of the CGM.

In future work, we will apply this model to cosmological simulations of
galaxies, and incorporate improved measurements of the observationally-derived
parameters constraining the total momentum budget ($p_0$) and time-scale
($t_{\rm FB}$), including their dependence on the galactic environment. With
this, we will be able to disentangle the effects of early feedback and the
cosmological assembly of galaxies in realistic environments, and examine how the
effects we have measured in isolated, Milky Way-like galaxies produce changes
across the cosmic lifetime of simulated galaxies.

\section*{Acknowledgements}
We thank Volker Springel for giving us access to {\sc Arepo}.  We thank Sam
Geen, Jindra Gensior, Sarah Jeffreson, Steve Longmore, and Lachlan Lancaster for valuable
conversations regarding this paper.  BWK, JMDK, and MC gratefully acknowledge
funding from the European Research Council (ERC) under the European Union's
Horizon 2020 research and innovation programme via the ERC Starting Grant
MUSTANG (grant agreement number 714907).  BWK acknowledges funding in the form
of a Postdoctoral Research Fellowship from the Alexander von Humboldt Stiftung.
JMDK and MC gratefully acknowledge funding from the German Research Foundation
(DFG) in the form of an Emmy Noether Research Group (grant number KR4801/1-1)
and the DFG Sachbeihilfe (grant number KR4801/2-1).
\section*{Data Availability Statement}
The simulation data used in this paper will be shared on reasonable request to
the corresponding author.

\bibliographystyle{mnras}
\bibliography{references}

\appendix
\section{Convergence Properties}
\label{convergence}
\begin{figure}
    \includegraphics[width=\hsize]{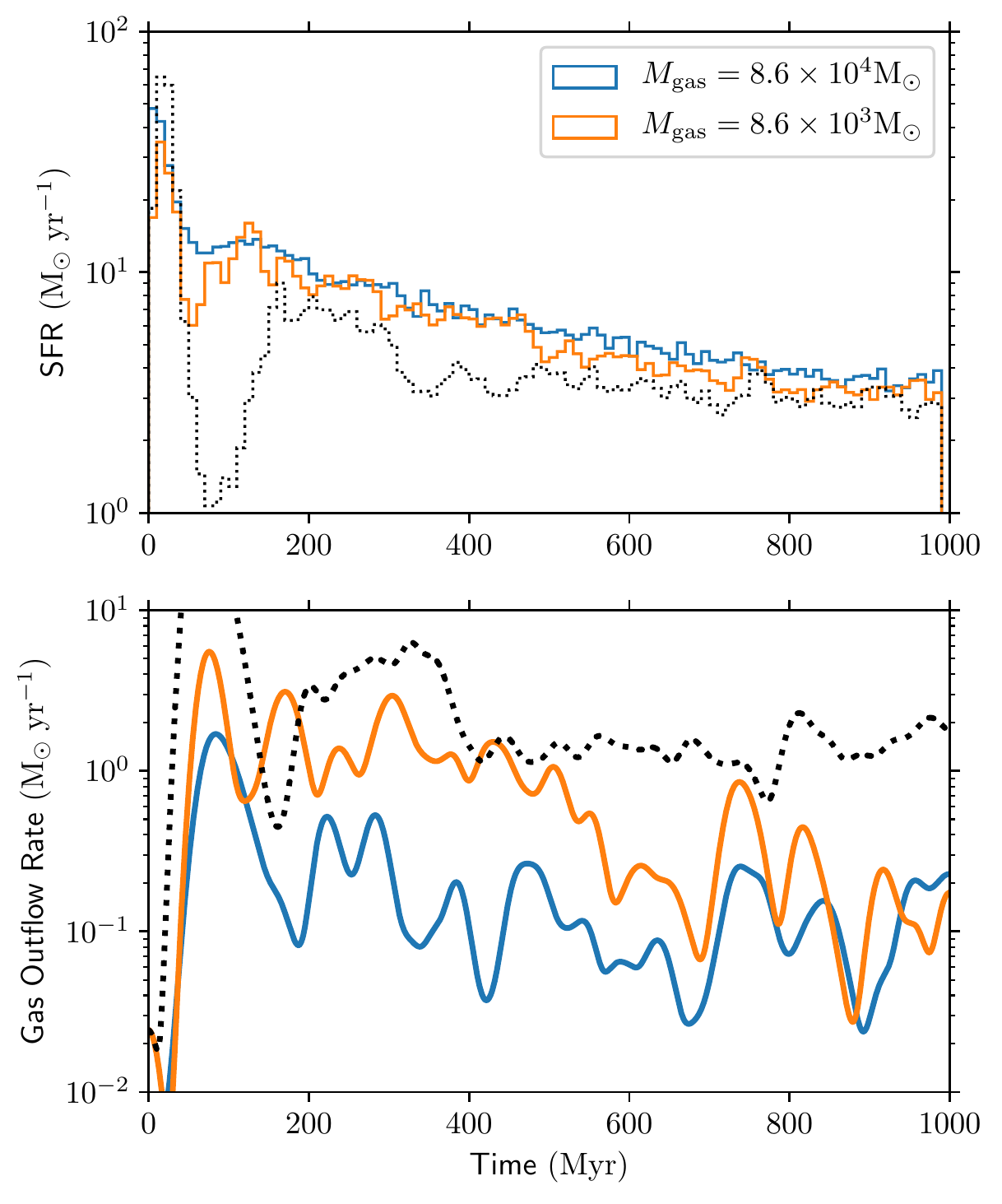}%
    \caption{Star formation and outflow rates for galaxies simulated with EMF at
    our fiducial (orange) and degraded (blue) resolutions.  The black dashed
    curve shows the results of our SN-only galaxy, run at the fiducial
    resolution. As can be seen, there is a slightly larger burst in at the start
    of the "settling" phase ($<50\Myr$) at the degraded resolution, as well as
    somewhat lower outflow rates during the first $\sim2$ orbits of the galaxy.
    For the final $\sim400\Myr$ of the galaxy's evolution, both the SFR and
    outflow rates are well converged.  As was shown previously, the settled SFR
    is only slightly increased with the addition of EMF, while the outflow rates
    are noticeably depressed.}
    \label{sfr_convergence}
\end{figure}
\begin{figure}
    \includegraphics[width=\hsize]{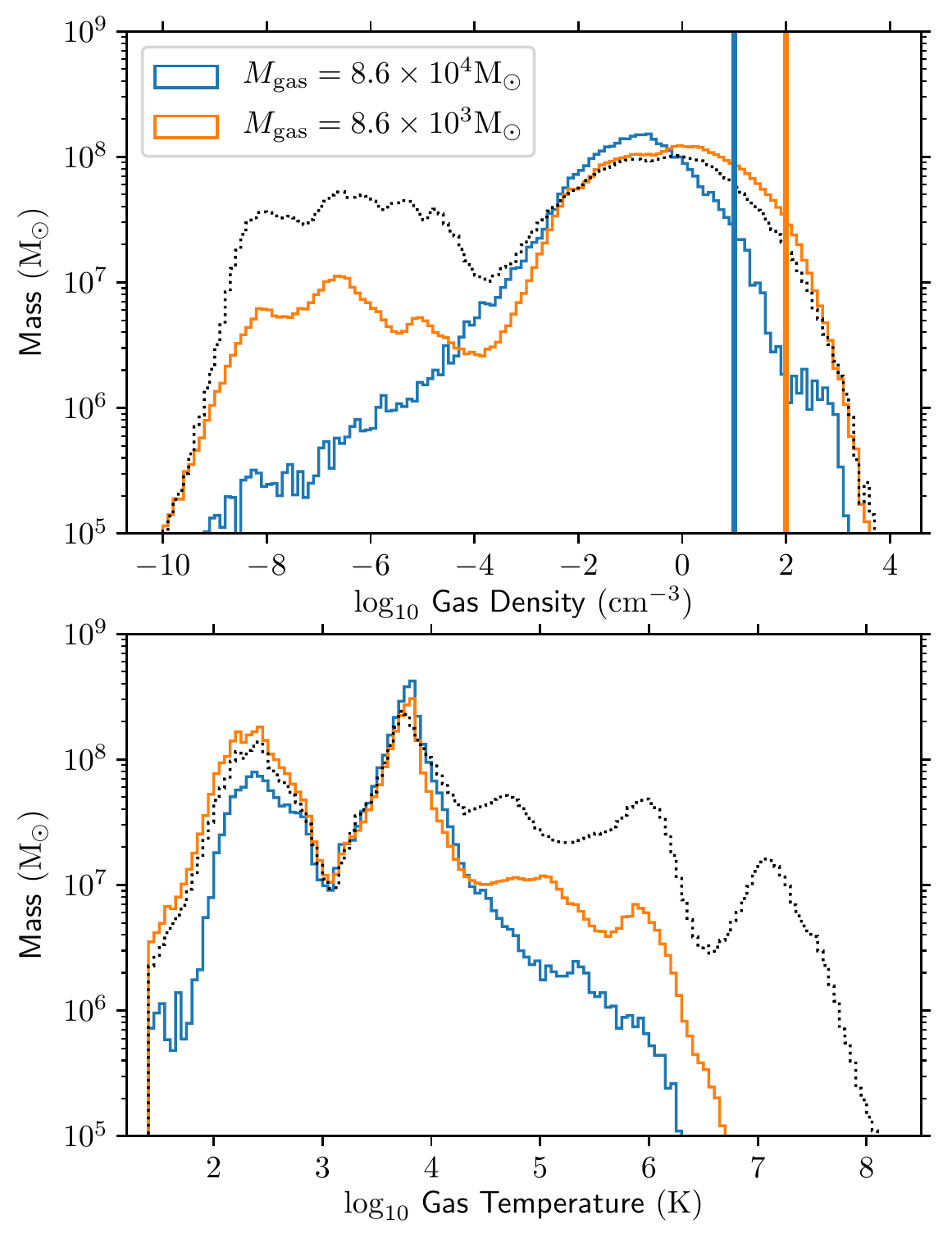}%
    \caption{Histograms of density (top panel) and temperature (bottom panel)
    for our fiducial and degraded resolution.  As in
    Figure~\ref{sfr_convergence}, the fiducial resolution is shown in orange,
    the degraded resolution in blue, and the SN-only result in dashed black.  In
    the density histogram, we have annotated the star formation density
    threshold as the two vertical lines, for the star formation threshold of
    $10\hcc$ and $100\hcc$ in the degraded and fiducial simulations
    respectively.  As can be seen, the degraded resolution produces slightly
    less cold, dense gas ($T<10^3$~K and $n>1\hcc$), likely due to the lower
    star formation threshold as well as the larger gas softening lengths.  We
    also see that the degraded resolution also produces less hot, diffuse gas
    ($T>10^5$~K and $n<10^{-3}\hcc$), suggesting that the mechanical SN
    algorithm is switching to momentum injection mode due to the lower
    resolution.  Despite this, both resolutions with EMF produce less hot,
    diffuse gas than the SN-only galaxy, and both have roughly equivalent gas in
    a warm-neutral phase ($T<10^5$~K).}
    \label{phase_convergence}
\end{figure}

The EMF model has been designed specifically for use in simulations that do not
resolve the detailed interior structure of molecular clouds and the interaction
of early feedback processes within these clouds.  The fiducial resolution used
in the isolated galaxy simulations we have presented here $(M_{\rm
gas}=8.6\times10^3\Msun)$ is comparable to existing high-resolution cosmological
zoom-in galaxy simulations \citep{Guedes2011,Sawala2016,Hopkins2018b,Font2020},
but is significantly higher than is achievable in large-volume cosmological
simulations \citep{Tremmel2017,Nelson2019}.  In order to ensure the impact of
our new models converge at the lower resolution required for cosmological
(rather than isolated) simulations, we have re-run the simulation using EMF with
the median parameters $p_0=377\kms$, $t_{\rm FB}=3.3\Myr$, and with $\alpha=1.0$
at a degraded resolution.  The degraded simulation has the same global disc
properties as the fiducial resolution, but with all particle masses increased by
a factor of 10, such that the degraded DM mass resolution is
$1.254\times10^7\Msun$, while the baryonic mass resolution is
$8.6\times10^4\Msun$.  We also increase the softening by a factor of two (to
$80\pc$), and decrease the minimum density threshold for star formation to
$1\hcc$ to account for the reduced ability to resolve dense gas at the degraded
resolution.

The SFR and outflow rates for this test are shown in
Figure~\ref{sfr_convergence}.  As can be seen from this Figure, the star
formation rate and outflow rates are reasonably converged, with some stochastic
variation between the fiducial and degraded resolution, especially for the
duration of their re-equilibriation after the start of the simulation (the first
$\sim600$~Myr).  Both resolutions show the same difference compared to the
SN-only run, with roughly comparable star formation rates, but outflow rates
lower by $\sim1$ dex.

We see a similar weak dependence on resolution for the gas temperature and
density distributions shown in Figure~\ref{phase_convergence}.  Again, we show
the results of galaxies simulated with EMF at the fiducial and degraded
resolution, compared to the SN-only case simulated at the fiducial resolution.
With degraded resolution, we see a reduction in the mass of the densest gas, in
part due to the lowered star formation density threshold (shown as vertical
lines in the top panel of Figure~\ref{phase_convergence}).  Interestingly, we
also see a reduction in very hot, diffuse gas, which is in line with what we
would expect from the slight reduction in gas outflow rates shown in
Figure~\ref{sfr_convergence}.  The differences we see between the two resolution
cases are well below the differences between including or omitting EMF.
Regardless of resolution, EMF produces significantly less hot, diffuse gas than
with SN feedback alone.
\section{Robustness to Stochasticity}
\label{stochasticity}
\begin{figure}
    \includegraphics[width=\hsize]{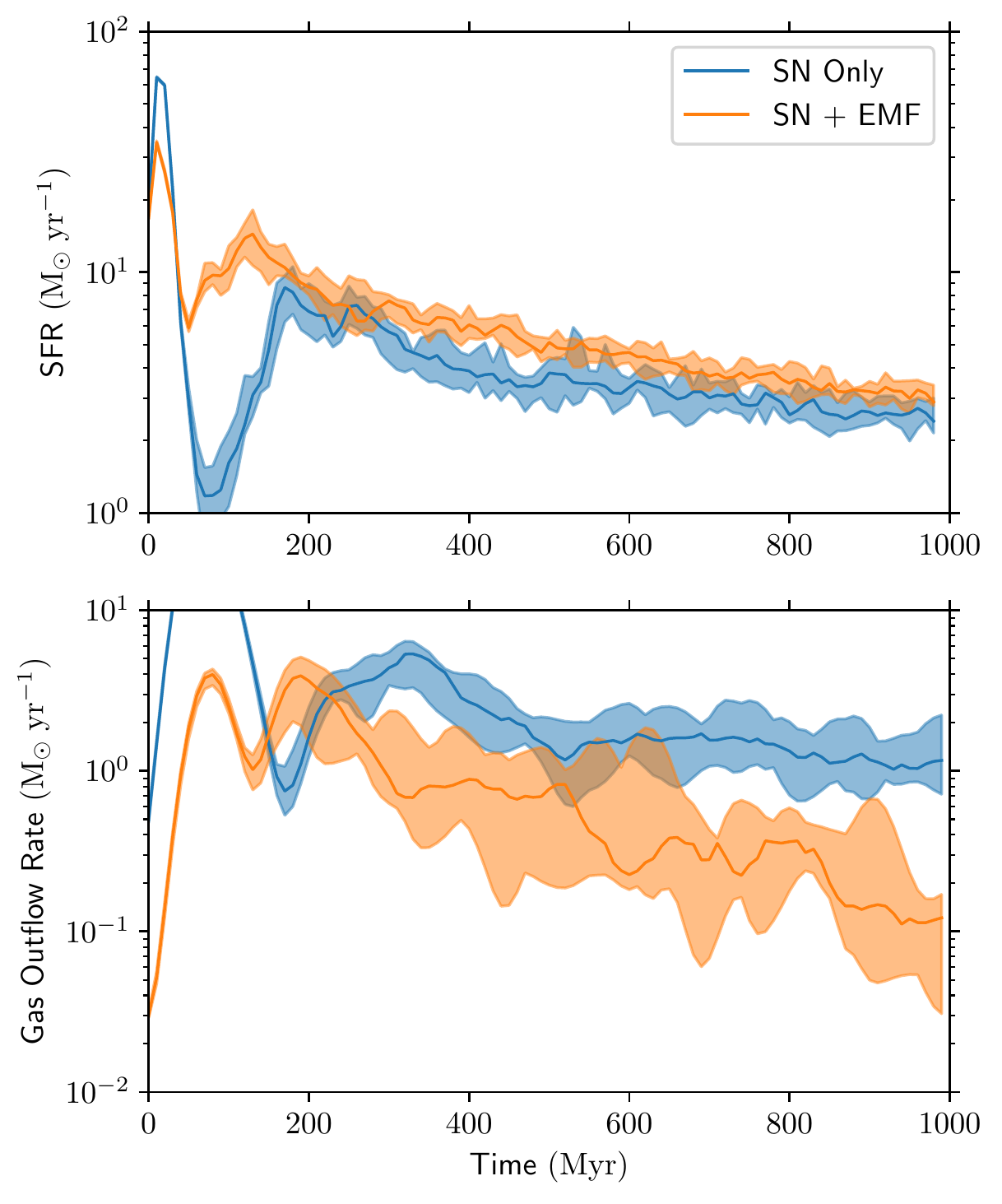}%
    \caption{Scatter in star formation (top panel) and outflow (bottom panel)
    rates for 8 separate runs with infinitesimal perturbations.  Blue shaded
    regions show the range of rates for SN-only runs, while orange shaded
    regions show the range for SN feedback together with EMF. The solid curve
    shows the median values.  As can be seen, the star formation rates for late
    times $>500\Myr$ are roughly within the scatter of each other.  Outflow
    rates, while showing larger stochasticity, still show the $\sim1$ dex
    difference in the median rates, consistent with the single runs we examined
    in the previous sections.}
    \label{stochasticity_sfr}
\end{figure}
\begin{figure}
    \includegraphics[width=\hsize]{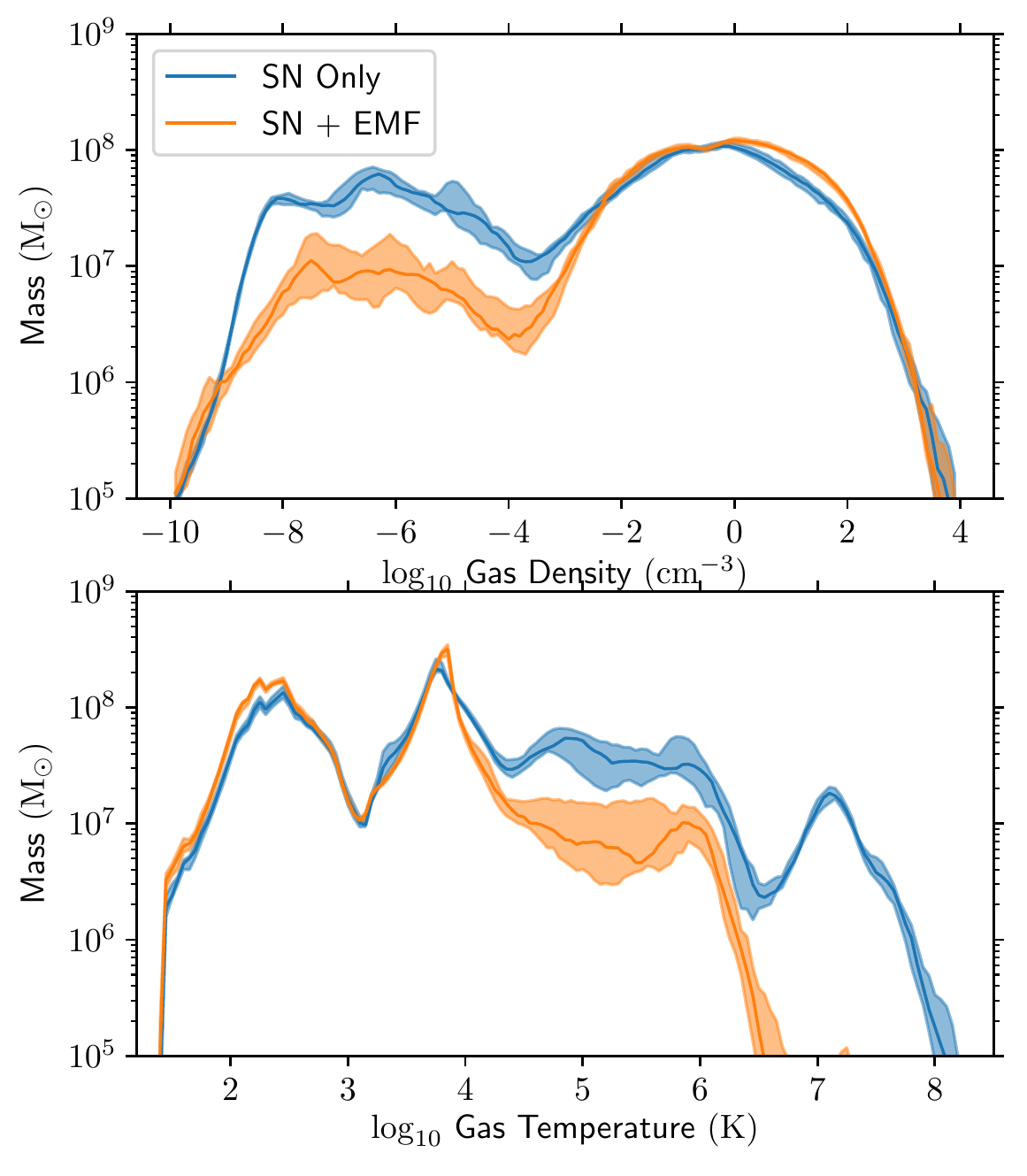}%
    \caption{Scatter in gas density (top panel) and temperature (bottom panel)
    PDFs for 8 perturbed simulations.  As in Figure~\ref{stochasticity_sfr},
    blue regions and lines show results for SN alone, while orange regions and
    lines show results from including EMF.  We can see that the differences
    between the SN-only runs and those which include EMF are well beyond the
    run-to-run scatter, primarily in the hottest, most diffuse gas.  As this gas
    is part of a relatively bursty outflow, we see that the scatter in gas above
    $10^4\K$ is considerably higher than for cooler neutral/ionized gas.}
    \label{stochasticity_phase}
\end{figure}

It has been demonstrated that galaxy evolution is a process that is, at some
level, chaotic and stochastic \citep{Keller2019,Genel2019}.  Small-scale
perturbations in initial conditions or numerical round-off can produce
variations in the overall star formation rate, chemical abundances, and
morphological properties of galaxies.  Here we demonstrate that the differences
we see between our simulations that include EMF and those with SN feedback only
are not explained by stochastic variation alone. To this end, we have
re-simulated both our $\alpha=1.0$ EMF simulation and the SN-only simulation 8
times, in order to estimate the run-to-run stochasticity in quantities we
measure.  Because {\sc Arepo} is fully deterministic, we have introduced a small
perturbation into gravity calculations by varying the opening angle $\theta$ for
gravity calculations (this parameter is given as {\sc ErrTolTheta} in {\sc
Arepo}) by $\mathcal{O}(10^{-6})$.  This introduces tiny differences in the
accelerations calculated by the gravity solver, which can then grow through the
mechanisms identified in \citep{Keller2019}.  As was shown previously, the
expansion exponent $\alpha$ has little impact of EMF, so for these tests we
restrict our EMF parameters to the fiducial values of $\alpha=1$, $p_0=375\kms$,
and $t_{\rm FB}=3.3\Myr$.

In Figure~\ref{stochasticity_sfr}, we show the star formation and outflow rates
for our stochasticity tests.  As in Figure~\ref{alpha_sfr}, the star formation
rates for the SN-only and EMF runs converge after an initial settling period,
while the outflow rates diverge to $\sim1$ dex differences in the median rates.
The scatter in outflow rates is much higher than the SFR.  \citet{Keller2019}
showed that temporal stochasticity is a reasonable proxy for run-to-run
stochasticity, and the outflow rates we found earlier in Figure~\ref{alpha_sfr}
show significant temporal variation (burstiness).  The larger scatter we see in
the outflow rates is consistent with this.  Despite the large scatter, the
median outflow rates for each feedback mechanism are consistently outside the
scatter of the other's run-to-run variation.  

Figure~\ref{stochasticity_phase} shows the variance in gas phase properties.
The differences we see between the SN-only runs and the EMF runs are again
significant beyond the run-to-run scatter.  The scatter in the dense, cool
$(<10^4\K)$ ISM shows very little stochastic variation, while the hot, diffuse
gas shows much larger scatter.  Interestingly, the largest scatter occurs in the
most thermally-unstable gas, with temperatures between $(10^4-10^5\K)$.  As the
peak of the solar-metallicity cooling rate curve $\Lambda(T)$ is near $10^5\K$,
this gas has short cooling times, and is the most transient phase of the gas
within the galaxy.  Gas colder than this is mostly at the equilibrium
temperature (where UV heating matches cooling), while gas hotter than this cools
slowly through adiabatic expansion.  The gas we see in the intermediate phase is
either moving to higher temperatures via SN heating or cooling back to
$T<10^4\K$.

\section{Impact of Star Formation Parameters}
\label{sfparams}
In the previous sections, we have restricted our analysis to variations in the
parameters and model choices used for stellar feedback.  It has been shown that
the star formation model, and the parameters used in that model, can also have
significant impact on the evolution of the ISM and the galaxy as a whole
\citep[e.g.][]{Hopkins2011,Benincasa2016,Semenov2021}.  In order to probe the
relative impact of star formation parameters relative to our new feedback model,
we re-simulate our fiducial EMF $\alpha=1.0$ galaxy while varying the star
formation efficiency per free-fall time $\epsilon_{\rm ff}$ and the star formation
density threshold $n_{\rm SF}$ by a factor of 2 above and below the values used for
the other simulations in this paper.

\begin{figure}
    \includegraphics[width=\hsize]{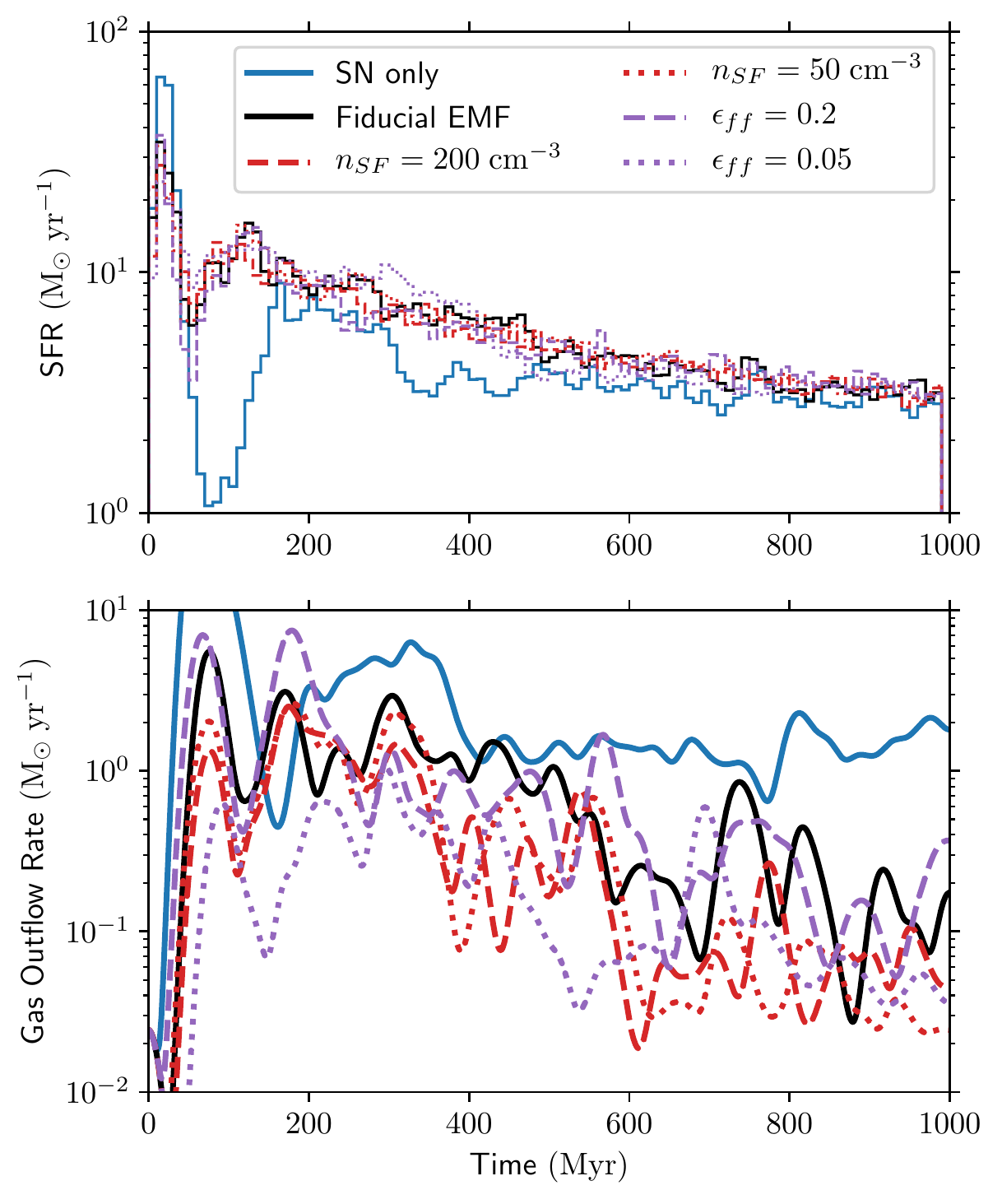}%
    \caption{Star formation (top panel) and outflows rates (bottom panel) for
    galaxies simulated with EMF and different parameters for the star formation
    efficiency per free-fall time $(\epsilon_{\rm ff})$ and star formation density
    threshold $(n_{SF})$.  As can be seen, varying either the star formation
    efficiency or the star formation threshold density does not change the
    overall star formation rate or the outflow rates beyond the stochasticity we
    show in Figure~\ref{stochasticity_sfr}.}
    \label{sfr_sfr}
\end{figure}

\begin{figure}
    \includegraphics[width=\hsize]{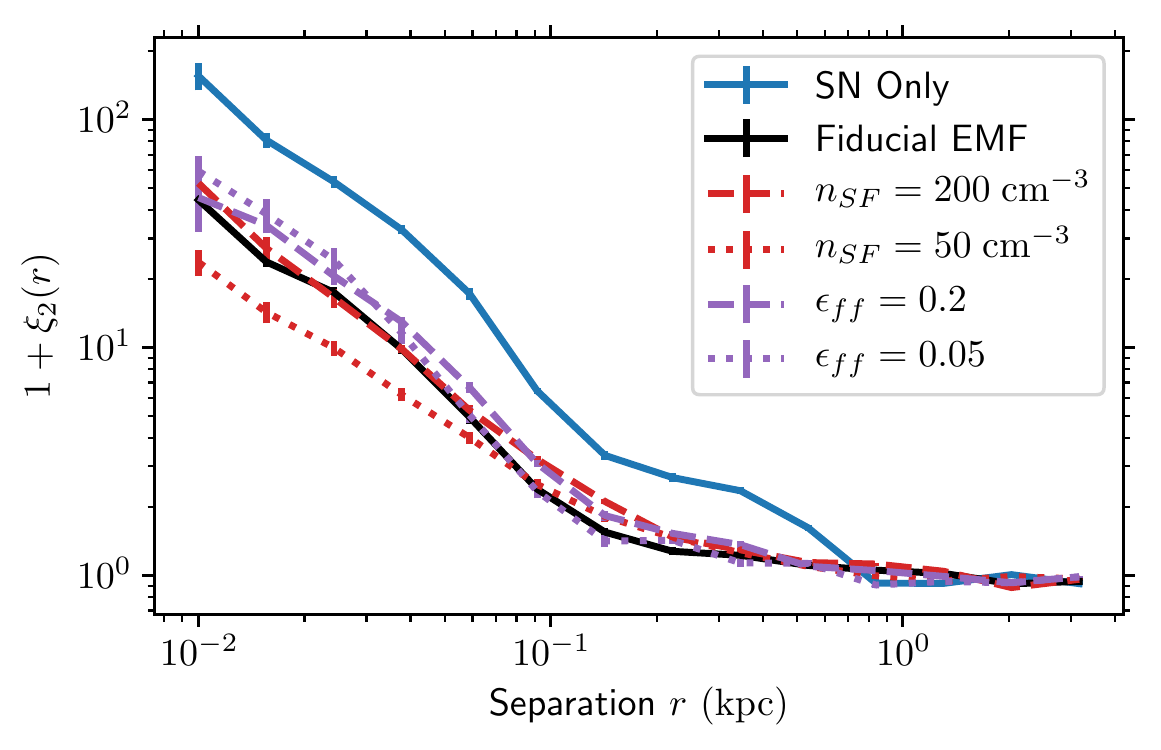}%
    \caption{Clustering of young stars as a function of star formation
    parameters $\epsilon_{\rm ff}$ and $n_{\rm SF}$ with EMF.  As can be seen, the
    reduced small-scale clustering of young stars is only slightly altered by
    changing the star formation parameters, with the largest impact being a
    slight reduction in the two-point correlation function below $100\pc$ with a
    lower density threshold $n_{\rm SF}=50\hcc$.}
    \label{sfr_clustering}
\end{figure}

\begin{figure}
    \includegraphics[width=\hsize]{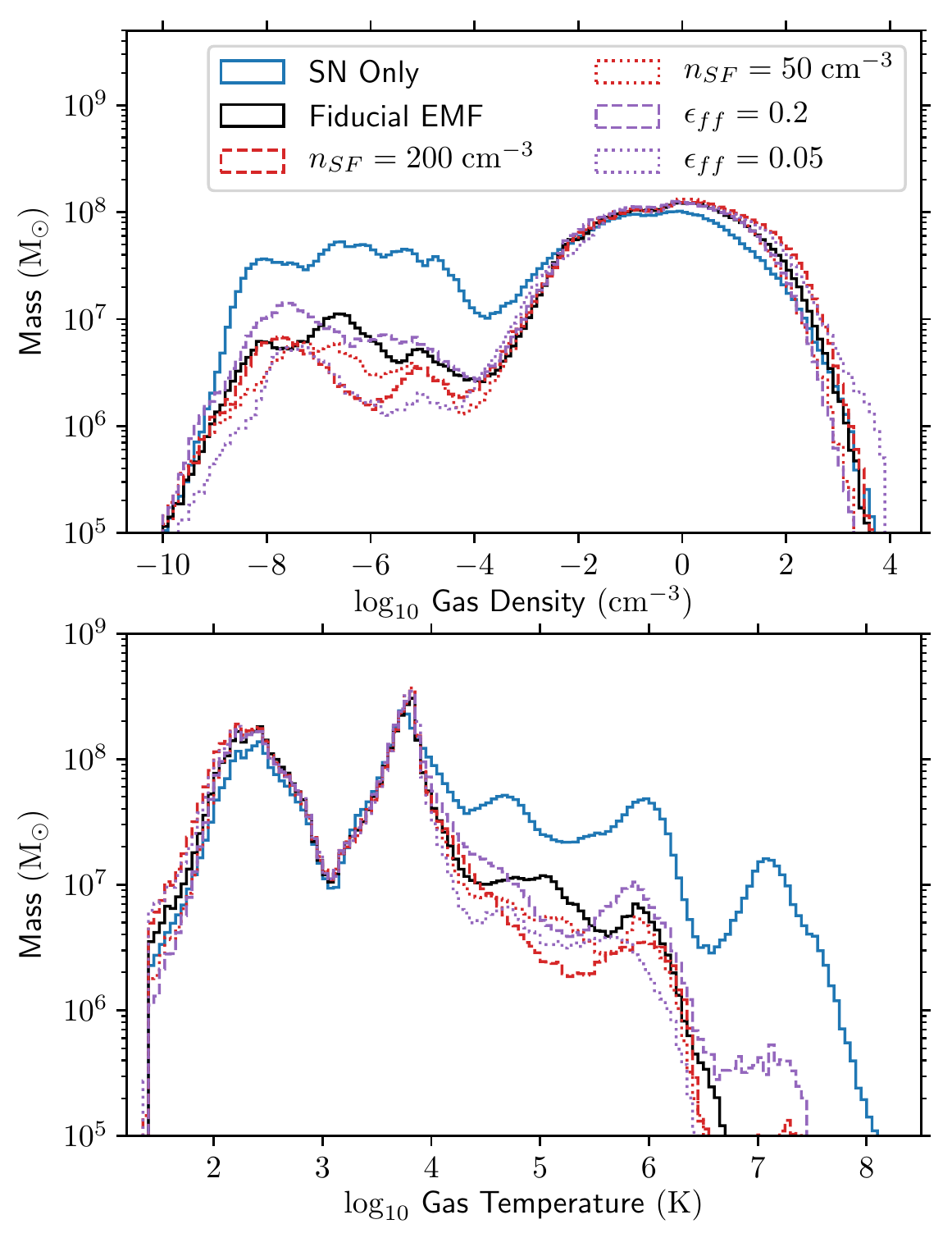}%
    \caption{Gas density (top panel) and temperature (bottom panel) PDFs for
    galaxies simulated with EMF and different star formation parameters.  As can
    be seen, the differences in both the gas temperature and density
    distributions for different star formation parameters are within the scatter
    shown in Figure~\ref{stochasticity_phase}.}
    \label{phase_sfr}
\end{figure}

In general, these parameter changes produce little impact on the overall evolution of
the galaxy.  As we show in Figure~\ref{sfr_sfr}, both the star formation and
outflow rates are within the the scatter produced by simple galactic
stochasticity shown in Figure~\ref{stochasticity_sfr}.  Clustering of young
stars are also only weakly altered by changing the star formation parameters, as
we show in Figure~\ref{sfr_clustering}.  Only in the case where we reduce the
star formation threshold density to $n_{\rm SF}=50\hcc$ do we see statistically
significant differences, with reduced clustering on scales below $100\pc$.  This
is mostly what we would expect, as reducing the threshold density for star
formation will allow star forming gas to begin forming stars earlier, when a
cloud has not collapsed to the smaller sizes that would be required to reach
higher densities.  The weak effect we see in the star formation properties also
translate to a weak dependence on the star formation parameters for gas
properties.  In Figure~\ref{phase_sfr}, we show the PDFs for gas density and
temperature.  As can be seen, both quantities show differences within the
stochastic scatter shown in Figure~\ref{stochasticity_phase}.  The increase in
very hot gas ($>10^7\K$) seen for the $\epsilon_{\rm ff}=0.2$ case is the result of a
late-time bursty outflow that occurs just before $t=1\Gyr$, and is not present
for the gas temperature PDF from $t=900\Myr$.

We show the changes to the star formation quantities measured by {\sc
Heisenberg} in Table~\ref{sf_heisenberg}. As this table shows, we see stronger
changes in both the cloud lifetimes $t_{\rm gas}$ and feedback timescale $t_{\rm
FB}$  when the star formation threshold density $n_{\rm SF}$ is changed, compared to
varying the sub-grid star formation efficiency $\epsilon_{\rm ff}$.  As we would
expect, lowering $\epsilon_{\rm ff}$ results in longer-lived clouds.  Lowering
$n_{\rm SF}$ pushes the effect in the opposite direction: we find shorter-lived
clouds, which are disrupted more quickly, and form a smaller fraction of their
mass into stars.  A higher star formation threshold requires gas to reach higher
densities to form stars: this results in clouds which take longer to begin star
formation (and therefore feedback).  These denser clouds will also be more
difficult to unbind through feedback, which results in a longer feedback
timescale.  The effects of sub-grid star formation parameters on the spatial
decorrelation of dense gas and young stars was explored in \citet{Semenov2021},
which also found that the tuning fork diagram is sensitive both to model choices
for feedback and star formation.  We leave a more comprehensive study of the
spatial decorrelation and cloud-scale star formation for star formation, rather
than feedback, model choice to a future study.

\begin{table}
\centering
    \begin{tabular}{|l|r|r|}
        Galaxy & $t_{\rm gas}$ (Myr) & $t_{\rm FB}$ (Myr) \\
        \hline
        \hline
        LMC (observations) & $11.1^{+1.6}_{-1.7} $ & $1.1^{+0.3}_{-0.2}$ \\
        PHANGS (observations) & $19.8^{+3.0}_{-2.0}$ & $3.31^{+0.83}_{-0.76}$ \\
        SN Only & $14.8^{+1.7}_{-1.5}$ & $3.3^{+0.7}_{-0.7}$ \\
        Fiducial EMF & $9.3^{+1.3}_{-0.8}$ & $1.4^{+0.6}_{-0.5}$ \\
        EMF $(n_{\rm SF}=200\hcc)$ & $20.4^{+2.2}_{-1.9}$ & $2.4^{+0.6}_{-0.6}$ \\
        EMF $(n_{\rm SF}=50\hcc)$ & $4.4^{+0.3}_{-0.4}$ & $1.0^{+0.2}_{-0.2}$ \\
        EMF $(\epsilon_{\rm ff}=0.2)$ & $6.6^{+0.9}_{-0.6}$ & $1.4^{+0.5}_{-0.6}$ \\
        EMF $(\epsilon_{\rm ff}=0.05)$ & $14.2^{+1.8}_{-1.6}$ & $1.1^{+0.5}_{-0.6}$ \\
        \hline
    \end{tabular}
    \caption{Time-scales and cloud-scale star formation efficiencies measured
    using {\sc Heisenberg} for observations of the LMC and EMF with different
    sub-grid star formation parameters.  As can be seen, increasing the density
    threshold for star formation increases $t_{\rm gas}$, and $t_{\rm FB}$.  The
    same effect occurs when decreasing the subgrid star formation efficiency.}
    \label{sf_heisenberg}
\end{table}

\label{lastpage}
\bsp

\end{document}